\documentclass[lettersize,journal]{IEEEtran}
\usepackage{amsmath,amsfonts}
\usepackage{amsmath}
\usepackage{amssymb}
\usepackage{amsthm}
\usepackage{algorithmic}
\usepackage{algorithm}
\usepackage{array}
\usepackage{enumerate}
\usepackage[caption=false,font=normalsize,labelfont=sf,textfont=sf]{subfig}
\usepackage{textcomp}
\usepackage{stfloats}
\usepackage{url}
\usepackage{verbatim}
\usepackage{graphicx}
\usepackage{epstopdf}
\usepackage{subfig}
\usepackage{cite}
\usepackage{balance}
\usepackage[numbers,sort&compress]{natbib}
\usepackage{lipsum,multicol}
\usepackage{textcomp,mathcomp}
\usepackage{xcolor}
\usepackage{makecell}
\usepackage{multirow}
\usepackage{booktabs}
\usepackage{gensymb}
\usepackage{flushend}
\newtheorem{problem}{Problem}
\theoremstyle{definition}
\newtheorem{remark}{Remark}
\begin{document}

\title{Scalable Optimal Power Management for Large-Scale Battery Energy Storage Systems}

\author{Amir Farakhor,~\IEEEmembership{Graduate Student Member,~IEEE}, Di Wu,~\IEEEmembership{Senior Member,~IEEE},
		Yebin Wang,~\IEEEmembership{Senior Member,~IEEE}, and Huazhen Fang,~\IEEEmembership{Member,~IEEE}
\thanks{This work was supported in part by the U.S. National Science Foundation under Award CMMI-1847651, and in part by the U.S. Department of Energy's Office of Electricity, Energy Storage Division, under Contract DE-AC05-76RL01830.}

\thanks{A. Farakhor and H. Fang (corresponding author) are with the Department of Mechanical Engineering, University of Kansas, Lawrence, KS, USA (Email: fang@ku.edu, a.farakhor@ku.edu).} 
\thanks{D. Wu is with the Pacific Northwest National Laboratory, Richland, WA, USA (Email: di.wu@pnnl.gov).}
\thanks{Y. Wang is with the Mitsubishi Electric Research Laboratories, Cambridge, MA, USA (Email: yebinwang@merl.com).}
}
\markboth{}%
{Shell \MakeLowercase{\textit{et al.}}: A Sample Article Using IEEEtran.cls for IEEE Journals}

\maketitle

\begin{abstract}
Large-scale battery energy storage systems (BESS) are helping transition  the world towards sustainability with their broad use, among others, in electrified transportation,  power grid, and renewables. However, optimal power management for them is often computationally formidable. To overcome this challenge, we develop a scalable approach in the paper. The proposed approach partitions the constituting cells of a large-scale BESS into clusters based on their state-of-charge (SoC), temperature, and internal resistance. Each cluster is characterized by a representative model that approximately captures its collective SoC and temperature dynamics, as well as its overall power losses in charging/discharging. Based on the clusters, we then formulate a problem of receding-horizon optimal power control to minimize the power losses while promoting SoC and temperature balancing. The cluster-based power optimization will decide the power quota for each cluster, and then every cluster will split the quota among the constituent cells. Since the number of clusters is much fewer than the number of cells, the proposed approach significantly reduces the computational costs, allowing optimal power management to scale up to large-scale BESS. Extensive simulations are performed to evaluate the proposed approach. The obtained results highlight a significant computational overhead reduction by more than 60\% for a small-scale and 98\% for a large-scale BESS compared to the conventional cell-level optimization. Experimental validation based on a 20-cell prototype further demonstrates its effectiveness and utility.
\end{abstract}


\section*{Nomenclature}
\addcontentsline{toc}{section}{Nomenclature}
\begin{IEEEdescription}[\IEEEusemathlabelsep\IEEEsetlabelwidth{$V_1,V_2,V_3$}]

\item[\textbf{Variables}]

\item[$v$] Cell terminal voltage
\item[$u$, $\bar{u}$] Cell open-circuit voltage (OCV), cluster OCV
\item[$i_L$, $\bar{i}_L$] Cell current, cluster current
\item[$P_b$, $\bar{P}_S$] Cell internal power, cluster internal power
\item[$P_l$, $\bar{P}_l$] Cell power loss, cluster power loss
\item[$P$] Cell output power
\item[$E$, $\bar{E}$] Cell energy, lumped cluster energy
\item[$P_{\textrm{out}}$] BESS output power demand
\item[$L$, $\bar{L}$] Total power losses of the cells, total power losses of the clusters
\item[$\xi$, $\bar{\xi}$] Intra- and inter-cluster slack variables
\item[$z$, $\bar{z}$] Intra- and inter-cluster optimization variables

\vspace{0.25cm}
\item[\textbf{Parameters}]

\item[$n$] Number of battery cells
\item[$n_S$] Number of cells within cluster $S$
\item[$k$] Number of clusters
\item[$m$, $\bar{m}$] Mass of a cell, lumped mass of a cluster
\item[$Q$, $\bar{Q}$] Cell capacity, lumped cluster capacity
\item[$q$, $\bar{q}$] SoC of a cell, lumped SoC of a cluster
\item[$q_{\textrm{avg}}$] Average SoC
\item[$\Delta q$, $\Delta \bar{q}$] SoC imbalance tolerance among cells, SoC imbalance tolerance among clusters
\item[$\Delta E$, $\Delta \bar{E}$] Energy balancing threshold among cells, energy balancing threshold among clusters
\item[$\alpha$] Intercept coefficient of the SoC/OCV line 
\item[$\beta$] Slope coefficient of the SoC/OCV line 
\item[$R$, $\bar{R}$] Cell internal resistance, lumped cluster internal resistance
\item[$R_C$] Resistance to capture the power losses of DC/DC converters
\item[$R_{\textrm{conv}}$, $\bar{R}_{\textrm{conv}}$] Cell convective thermal resistance, cluster convective thermal resistance
\item[$A$] Cell external surface.
\item[$h$] Conductive heat transfer coefficient between the cell's surface and the environment
\item[$C_{\textrm{th}}$, $\bar{C}_{\textrm{th}}$] Cell thermal capacitance, lumped cluster thermal capacitance
\item[$T$, $\bar{T}$] Cell temperature, lumped cluster temperature
\item[$T_{\textrm{avg}}$, $\bar{T}_{\textrm{avg}}$] Average cells' temperature, average clusters' temperature
\item[$T_{\textrm{env}}$] Environmental temperature
\item[$\Delta T$, $\Delta \bar{T}$] Temperature imbalance tolerance within cells and clusters
\item[$\lambda$, $\bar{\lambda}$] Penalty weight for the intra- and inter-cluster optimization
\item[$\Delta t$] Sampling time
\item[$H$] Optimization horizon
\end{IEEEdescription}

\section{Introduction}
\IEEEPARstart{B}{attery} energy storage systems (BESS) have emerged as an enabler for various applications ranging from  electric vehicles (passenger cars, semi trucks, etc.),   electric aircraft, smart grid, and renewable facilities~\cite{TTE-ZW-2021,2022-ITEC-SM,2020-arXiv-BA,TTE-LY-2022}. Whether small or large in size, BESS need power management strategies to ensure their safe and proper operation, which provide various functions including charging/discharging control, cell balancing, and power loss minimization. While some simplistic approaches have gained wide use, there is a growing demand for more sophisticated optimal power management to maximize the performance and fully utilize the capabilities of BESS \cite{TTE-WY-2020,2019-EnergyResearch-SM}. However, optimal power management faces the challenge of high computational expenses, due to the use of numerical optimization. The challenge is especially intense for large-scale BESS comprising great numbers of cells. We broadly define large-scale BESS as those that offer high energy or power capacity and that comprise many constituent entities like cells or modules, with examples including grid- and vehicle-scale BESS. Optimal power management for them will involve large numbers of decision variables as well as complex high-dimensional optimization landscapes. Despite an increasing body of study on optimal BESS power management, only limited effort has been dedicated to overcoming the computational bottleneck in optimization.

\subsection{Literature Review}

Early optimal power management strategies in the literature mainly focused on cell balancing for BESS. The studies in \cite{2013-ECC-PM, 2015-ITEC-RG} use linear programming to analyze and evaluate the performance of different balancing circuit topologies for battery packs. However, the balancing considers only the SoC while neglecting some other important factors like temperature and internal resistance, and its computation can be heavy for large battery packs.

The use of power electronic converters in BESS circuit structures has enabled cell-level bidirectional power control. Leveraging this capability, the study in \cite{2014-VPPC-BJ} pursues optimal power management to achieve SoC/temperature balancing and terminal voltage regulation. The work in \cite{2016-TSTE-PC} further aims to minimize the total power losses of a BESS while making the cells satisfy safety and balancing requirements. It further convexifies the non-convex power optimization problem using a technique in \cite{2012-IFAC-NM} for the benefit of computation. The idea of convexified power optimization also finds success in a hybrid BESS consisting of battery cells and supercapacitors \cite{2019-TVT-CR} and a reconfigurable BESS \cite{TTE-FA-2022}.

Looking back, the existing optimal power management strategies rely on numerical optimization methods, and the need for computational resources increases fast with the number of cells of the concerned BESS. The literature has explored two methods to alleviate the computational burden, namely, hierarchical control and distributed control. A hierarchical model predictive control is proposed in \cite{2021-TCST-CR} with different time scales and model complexities. The framework considers both charge and temperature imbalances while minimizing the total power losses. The study in \cite{TCST-AF-2017} attempts to decompose an optimal power management task into separate voltage and balancing control subtasks. These methods can reduce the computation to a certain extent but still lack scalability for large-scale BESS.

Distributed control represents another means to accelerate large-scale BESS power management, which distributes computation among different computing units for higher efficiency \cite{2016-TPEL-MT,2020-TII-OQ}. The work in \cite{2016-TPEL-MT} employs a dynamic average consensus protocol to achieve SoC balancing for a reconfigurable BESS. An average consensus protocol is similarly used in \cite{2018-TSTE-OQ} to balance the SoC of serially connected BESS. Despite their fast computation, the average consensus-based power management strategies only focus on cell balancing, mostly in terms of SoC, without consideration of optimality in some metrics such as balancing time and power loss. The study in \cite{ACC-FA-2023} proposes a distributed approach for cell balancing and power loss minimization for large-scale BESS. However, it requires the use of distributed computing units and communication networks, thus increasing the complexity of implementation.

The notion of optimal power management extends to control of distributed energy resources, which often appears as the problem of optimal power dispatch among the resources. For instance, the studies in \cite{ITEC-DG-2023, EnergyStorage-SN-2023, TTE-WS-2023} formulates different optimal power management problems and solutions for shipboard microgrid and all-electric ships. Optimal power management also finds applications in the domain of hybrid electric vehicles \cite{EnergyConversion-SF-2023}. In these applications, the computational costs is also  a topic of concern and interest.

\subsection{Statement of Contributions}

\begin{figure}[!t]
\centering
\includegraphics[width=8cm]{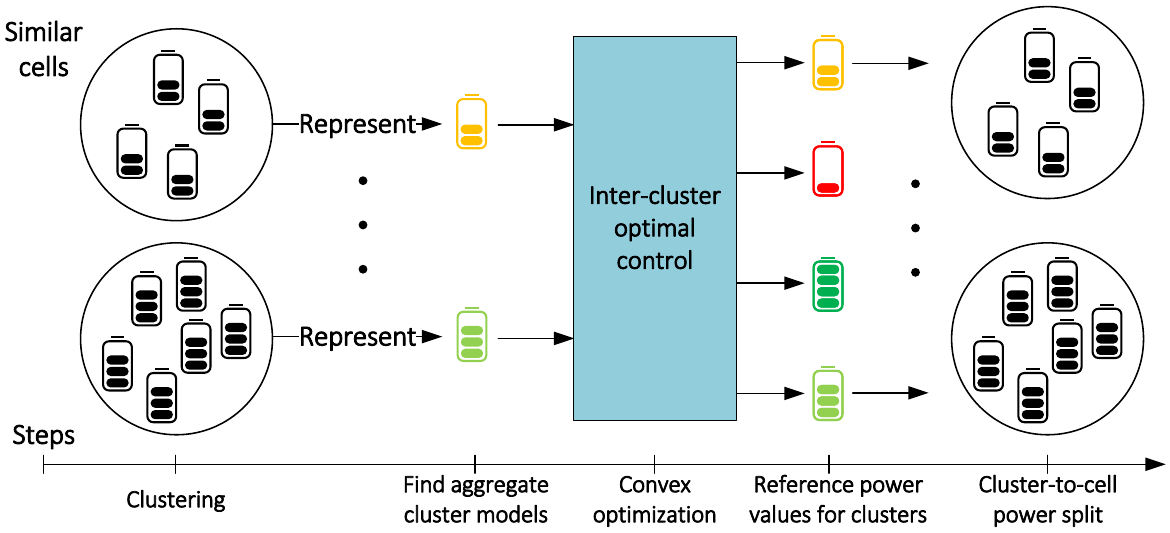}
\caption{The proposed clustering-based power management approach.}
\label{Pipeline}
\end{figure}

Departing from the literature, this paper proposes clustering-based optimization to achieve computationally efficient and scalable optimal power management for large-scale BESS. The key notion lies in grouping the cells of a BESS into clusters based on their characteristics, then performing inter-cluster power optimization, and finally running intra-cluster power allocation for individual cells. Centering around the notion, our main contribution are as follows. We develop a systematic design of a clustering-based power management approach. The approach, which is illustrated in Fig~\ref{Pipeline}, includes the following crucial elements.
\begin{itemize}
	\item We leverage the $k$-means clustering method to partition the cells into clusters according to their SoC, temperature, and internal resistance. The cells within a cluster will share similar characteristics while differing distinctly from those in another cluster.
	\item We develop a representative electro-thermal model for each cluster to capture the cluster’s aggregated dynamics in SoC, temperature, and power loss in operation. Then, using the cluster models, we formulate and solve an optimal power management problem.
	\item Following the cluster-level optimization, we design three power split schemes to determine the power to be assigned for the cells within each cluster.
\end{itemize}
   
The proposed approach is hierarchical---it breaks down a large-scale optimal power management management problem into two levels, inter-cluster power optimization and intra-cluster power allocation. The subproblems at each level are considerably smaller in size and simpler in structure to allow fast computation, and the hierarchical design readily scales up to large BESS. Meanwhile, the clustering based on the cells' characteristics upholds the fidelity of the proposed approach in solving the original problem, thus ensuring the overall performance in power management. 

We develop a 20-cell battery pack and conduct a series of experiments, along with extensive simulations, to validate and assess the performance of the proposed approach. The results demonstrate the effectiveness of the proposed optimal power management approach in significantly reducing the computation while maintaining good accuracy.

\subsection{Organization}

The rest of the paper is organized as follows. Section II describes the circuit structure of the considered large-scale BESS and the corresponding optimal power management problem. Section III presents the cell clustering approach, shows how to develop a representative electro-thermal model for a cluster, and implements an inter-cluster power optimization problem for the clusters. We then derive three schemes to handle cluster-to-cell power split. In Sections IV and V, the simulation and experimental results demonstrate the effectiveness of the proposed scalable optimal power management approach. Finally, Section VI concludes the paper with final remarks.

\section{Overview of Optimal Power Management}
In this section, we first introduce the circuit structure of a large-scale BESS, and then present the overarching control objectives for the large-scale BESS optimal power management.
\subsection{Circuit Structure}
\begin{figure}[!t]
\centering
\includegraphics[width=8.5cm]{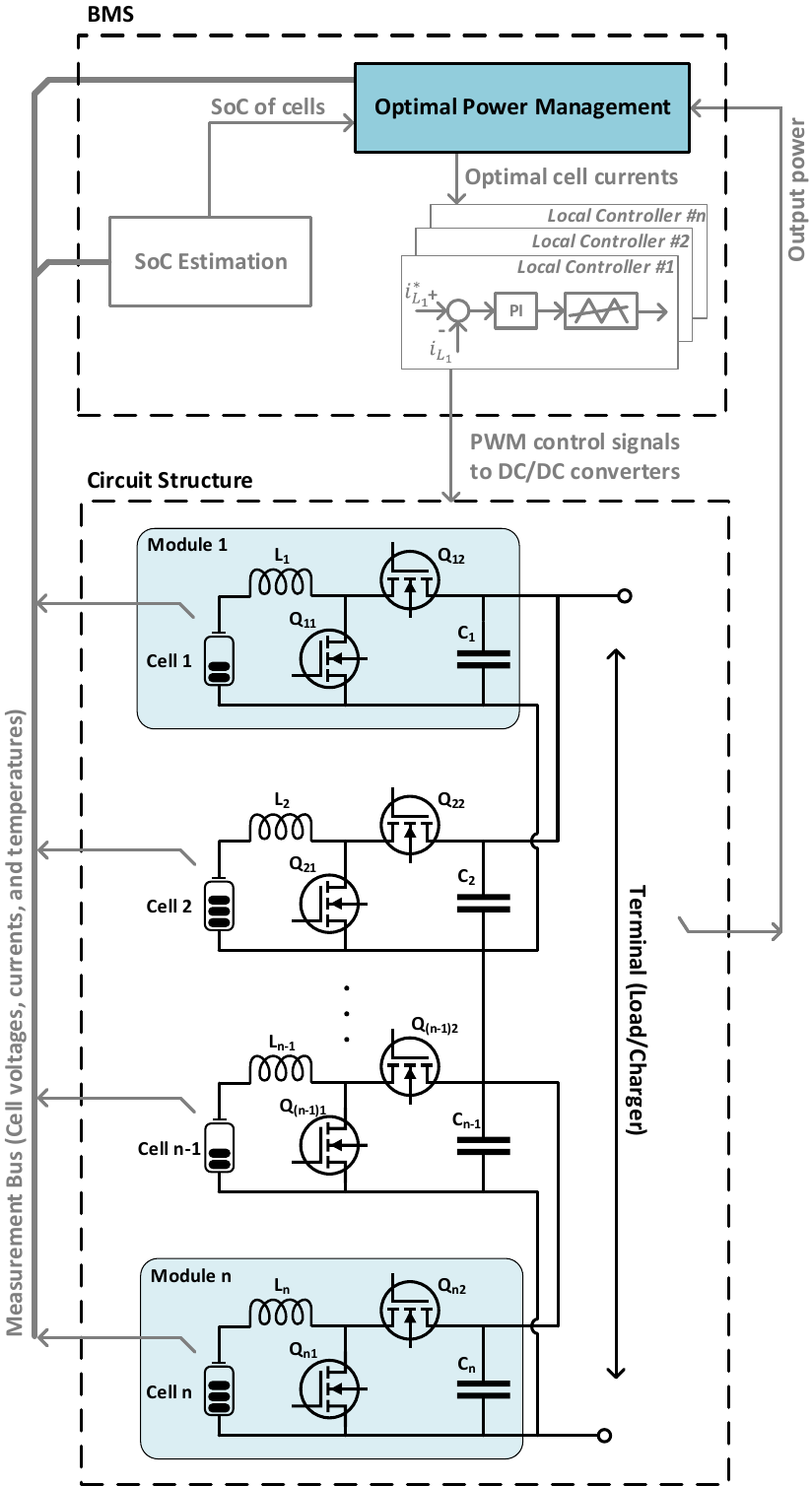}
\caption{The circuit structure of a large-scale BESS.}
\label{CircuitStructure}
\end{figure}

Fig.~\ref{CircuitStructure} depicts the circuit structure of a large-scale BESS. The BESS comprises $n$ modules, each consisting of a cell and a DC/DC converter. The modules are configured arbitrarily in series, parallel, or a mix of both to meet the output power, capacity, or voltage requirements. For the purpose of illustration, modules 1 and 2 are connected in parallel, and then connected serially with the other modules. The circuit structure is taken from our earlier study in \cite{2021-IECON-FA, TTE-FA-2022}, in which the module connections are reconfigurable via power switches. Here, we consider hardwired connections among the modules so as to focus on the optimal power management design. Here, we use synchronous DC/DC converters, but other types of bidirectional DC/DC converter topologies are allowed in the circuit structure. The DC/DC converter consists of an inductor, a capacitor, and two power switches. They allow controlled bidirectional power flow through cells in charging/discharging. Because of the DC/DC converters, one can independently regulate each cell's charging/discharging power. 

The cell-level power control capability of the circuit structure brings about several system-level advantages. First, this capability can be leveraged to promote the balanced use of the cells in terms of their SoC and temperatures. Second, unlike conventional structures, the circuit structure requires no external power electronic devices to regulate its output voltage---the embedded DC/DC converters can adjust the modules' output voltages to supply the load adequately. Finally, the circuit structure can accommodate heterogeneities among the constituent cells, charging or discharging them based on their individual conditions. In an extreme case, one can even leverage it to construct a large-scale BESS using cells from different manufacturers with different internal characteristics. It is worth mentioning that the circuit structure can be extended to modules or battery packs rather than cells. In this case, each battery pack will be equipped with a DC/DC converter to allow regulated charging and discharging.

Next, we define the control objectives in the optimal power management of this BESS circuit structure to fully take advantage of its capability.
\subsection{Control Objectives}
Fig.~\ref{CircuitStructure} also illustrates how the circuit structure interacts with the battery management system (BMS). The BMS comprises two types of controllers. At the higher level, the optimal power management block collects the real-time measurements of the cells and computes the optimal power allocation among them; at the lower level, the local controllers generate control signals for the DC/DC converters to regulate the cells' charging/discharging currents. This control architecture decouples power management and local power control, and this paper focuses only on the study of the former problem, with mature technologies available for the latter.

As discussed above in Section II.A, the circuit structure of the BESS allows cell-level power control, making it possible to achieve various functions, such as cell balancing and power loss minimization. To leverage this capability, we consider the following optimal power management problem in the paper:
\begin{problem}
Find the reference values for the cells' charging/discharging power to minimize the system-wide power losses while ensuring the cells to comply with the physical, safety, and balancing constraints and supplying the demanded output power.
\end{problem}

Problems of a similar kind have attracted various studies \cite{2021-TCST-CR,2019-TVT-CR,TTE-FA-2022} for different circuit structures. Existing studies generally adopt numerical optimization frameworks. Their computational complexity depends on the number of optimization variables, causing extremely heavy computational burdens for large-scale BESS. To tackle the challenge, this paper proposes a scalable optimal power management approach in the sequel.
\section{The Proposed Scalable Optimal Power Management}
This section presents the proposed scalable optimal power management approach, elaborating its four main elements: cell clustering, cluster model development, inter-cluster power optimization, and intra-cluster power split. To begin with, we partition the cells into clusters. We then introduce a representative electro-thermal model for the clusters. We formulate the inter-cluster optimal power management. Finally, we propose three schemes to split the power quota of clusters among their constituent cells.
\subsection{Clustering}
\begin{figure}[!t]
\centering
\includegraphics[trim={0 0 0 1.5cm},clip, width=8cm]{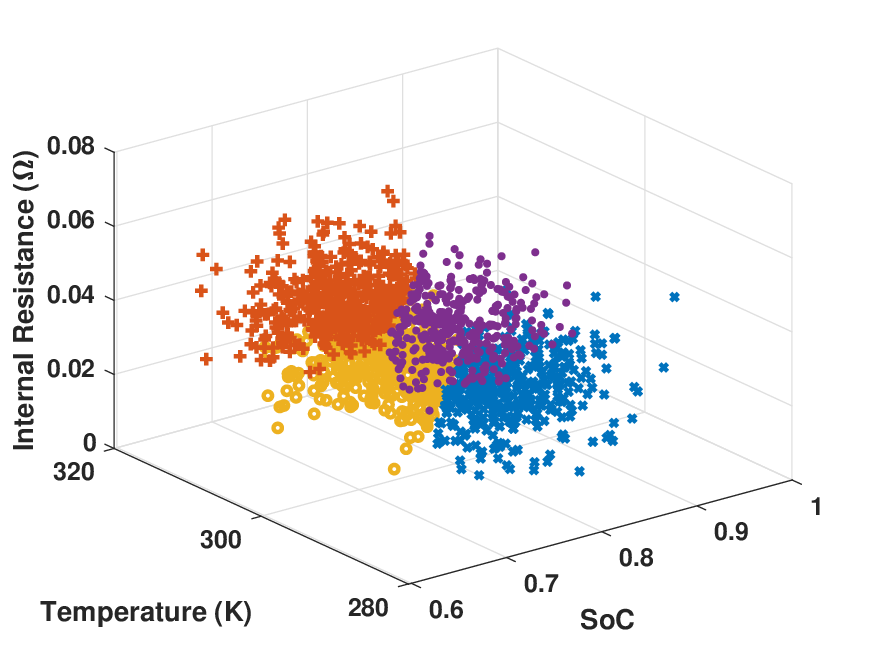}
\caption{Illustration of cell clustering for a large-scale BESS.}
\label{CellClustering} 
\end{figure}

Consider the BESS shown in Fig.~\ref{CircuitStructure} comprising $n$ cells, where $n$ is a large number. As is common in practice, the cells are heterogeneous in SoC, temperature, and internal resistance. An important objective in power management is to overcome the heterogeneity, or in other words, to balance the use of the cells via optimizing the power allocation among them based on their conditions. Running optimization over many heterogeneous cells, however, is computationally expensive. To treat this issue, we can group the cells into clusters. The clusters are much fewer in number, and every cluster includes cells with similar characteristics. We are then able to deal with cluster-level power optimization for computational benefits. 

To this end, we leverage the $k$-means clustering method to partition the $n$ cells into $k$ $(k\ll n)$ clusters $S_1,...,S_k$ \cite{PR-AL-2003}. This method is a favorable choice here because of its computational efficiency and capability to handle large datasets, even though alternative clustering techniques, e.g., density-based spatial clustering, are also applicable \cite{InfoSystems-LD-2007}. For cell $i$, its condition is characterized by the tuple $x_i=\{q_i,T_i,R_i\}$ for $i=1,2,...,n$, where $q_i$, $T_i$, and $R_i$ are the SoC, temperature and internal resistance, respectively. The clustering problem can be translated into the following optimization problem:
\begin{equation}
\begin{split}
&\min_{r_{ij}, c_j} \quad \sum_{i=1}^n \sum_{j=1}^k r_{ij} \left\| x_i-c_j \right\|^2,  \\
& \textrm{s.t.} \quad \quad \sum_{j=1}^k r_{ij} = 1 \quad \forall i=1,...,n,  \\
\end{split}
\label{Clustering}
\end{equation}
where $c_j$ is the centroid of cluster $S_j$, and $r_{ij}\in\{0,1\}$ with $r_{ij}=0$ if $x_i\notin S_j$ and $r_{ij}=1$ if $x_i\in S_j$. The problem in \eqref{Clustering} is NP-hard, often defying closed-form solution. Many heuristic algorithms have been proposed in the literature to solve it. Here, we use the naive $k$-means algorithm because of its effectiveness and efficiency. The algorithm starts with an arbitrary set of centroids $c_1(0),...,c_k(0)$ and follows an alternate two-step procedure \cite{IEEEAccess-SK-2020}. First, $x_i$ for $i=1,2,...,n$ are each assigned to its nearest centroid at the $\ell$-th iteration, i.e.,
\begin{equation}
r_{ik}(\ell)=
\begin{cases}
	1 \quad \textrm{if} \; k=\textrm{argmin}_j \left\|x_i-c_j(\ell)\right\|^2    \\
	0 \quad \textrm{else}
\end{cases}.
\end{equation}
Then, the centroids are refined and updated as follows:
\begin{equation}
c_j(\ell+1) = \frac{\sum_{i=1}^n r_{ij}(\ell)x_i}{\sum_{i=1}^n r_{ij}(\ell)}.
\end{equation}
The $k$-means algorithm iterates these two steps until convergence when the cluster assignments stop changing. Note that the algorithm requires to pre-specify the number of clusters $k$. Clearly, the more diverse the cells, the more clusters are needed to effectively categorize them. There are some useful techniques, e.g., elbow curve \cite{2022-AS-OA} or gap statistics \cite{2001-StatisticalSociety-TR}, to determine the optimal number of clusters, and gap statistics is the choice in this paper.

Fig.~\ref{CellClustering} illustrates the clustering of a 400-cell BESS for $k=4$. The clusters are labeled in different colors, and the cells within a cluster are closer to each other in terms of SoC, temperature, and internal resistance. Next, we will develop a representative model to capture the virtual collective dynamics of a cluster. 
\subsection{Electro-thermal Modeling}
We first introduce a cell-level electro-thermal model and then aggregate the cells to formulate a cluster-level model.
\subsubsection{Electrical Modeling}
\begin{figure}[!t]
    \centering
    \subfloat[\centering ]{{\includegraphics[width=4.9cm]{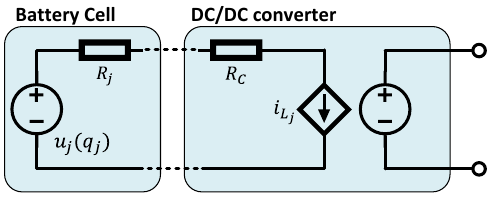} }}
    \subfloat[\centering ]{{\includegraphics[width=3.9cm]{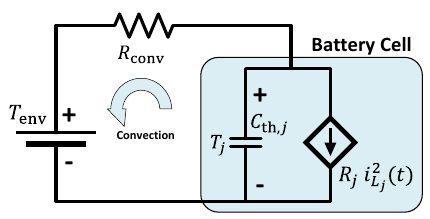} }}
    \caption{The cell-level electro-thermal model. (a) The electrical model of the module $j$. (b) The thermal model of the cell $j$.}
    \label{CellLevelModel}
\end{figure}

The electrical model of module $j$ is shown in Fig.~\ref{CellLevelModel} (a). We use the Rint Model to describe the electrical dynamics of cell $j$, which consists of an open-circuit voltage (OCV) and a series internal resistor \cite{2011-Energies-HH}. The Rint model offers a simple yet accurate enough representation of the cells. The model's governing equations are as follows:
\begin{subequations}
	\begin{align}
		\dot{q}_j(t)&=-\frac{1}{Q_j}i_{L_j}(t),\\
		v_j(t)&=u_j(q_j(t))-R_ji_{L_j}(t),
	\end{align}
\end{subequations}
where $Q_j$, $u_j$, $v_j$, and $i_{L_j}$ are the cell's capacity, OCV, terminal voltage, and charging/discharging current, respectively. Following \cite{TTE-FA-2022}, this paper assumes a piecewise linear approximation for the cell's SoC/OCV relationship as follows:
\begin{equation}
	u_j(q_j(t)) = \alpha_j^i(q_j(t)) + \beta_j^i(q_j(t))q_j(t),
\end{equation}
where $\alpha_j^i$ and $\beta_j^i$ are the intercept and slope coefficients of the $i$-th line segment. The cell's charging/discharging power can be expressed as
\begin{equation}
	P_{b_j}=u_j(q_j(t))i_{L_j}(t).
\end{equation}
Further, we idealize the DC/DC converter as a DC transformer with a series resistance $R_C$ to capture its power loss. The output power of the module is then given by
\begin{equation}
	P_j(t) = u_j(q_j(t))i_{L_j}(t) - (R_j+R_C)i_{L_j}^2(t),
\end{equation} 
where $R_ji_{L_j}^2(t)$ and $R_Ci_{L_j}^2(t)$ represent the power losses on the cell and converter, respectively. Next, we extend this cell-level model to the cluster level. 

\begin{figure}[!t]
\centering
\includegraphics[trim={0 0 0 0},clip, width=7cm]{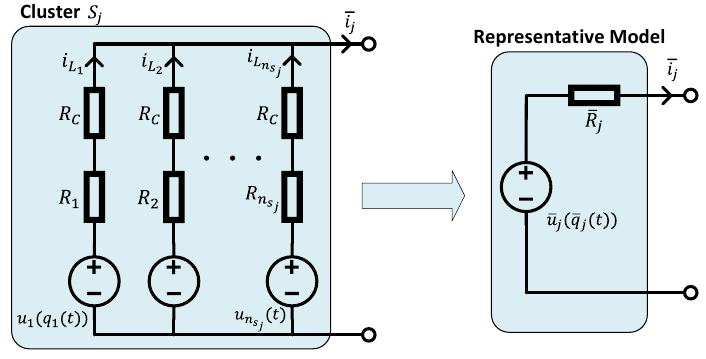}
\caption{Representative cluster-level electrical model.}
\label{ClusterLevelElectrical}
\end{figure}

Consider cluster $S_j$ with $n_{S_j}$ cells numbered from 1 to $n_{S_j}$, with $\sum_{j=1}^k n_{S_j}=n$. We intend to aggregate the cell models to derive a cluster model for $S_j$. Here, $S_j$ can be viewed as a virtual module, though the cells within it may not share hardwired connection. Then, we can assume the constituent cells to be connected virtually either in parallel or series. We choose virtual parallel connection here to derive the aggregate model for $S_j$. Fig.~\ref{ClusterLevelElectrical} illustrates the idea of lumping the cell-level models into a single model for cluster $S_j$.

For cluster $S_j$, the aggregated capacity $\bar{Q}_j$ and applied current $\bar{i}_{L_j}$ can be expressed as follows:
\begin{equation}
	\bar{Q}_j = \sum_{i=1}^{n_{S_j}} Q_i, \quad \bar{i}_{L_j} = \sum_{i=1}^{n_{S_j}} i_{L_i}.
	\label{AggregateCapacity}
\end{equation}
The cluster's SoC $\bar{q}_j$ is governed by
\begin{equation}
	\dot{\bar{q}}_j(t) = -\frac{1}{\bar{Q}_j}\bar{i}_{L_j}(t).
	\label{ClusterLevelSoCDynamic}
\end{equation}
Further, the cluster's internal resistance $\bar{R}_j$ and OCV $\bar{u}_j$ are as follows:
\begin{subequations}
	\begin{align}
		\bar{R}_j &= \frac{R_{n_{S_j}}+R_C}{1+\sum_{i=1}^{n_{S_j}-1}\frac{R_{n_{S_j}}+R_C}{R_i+R_C}},\\
		\bar{u}_j &= \frac{1}{1+\sum_{i=1}^{n_{S_j}-1}\frac{R_{n_{S_j}}+R_C}{R_i+R_C}}\left(\sum_{i=1}^{n_{S_j}}\frac{R_{n_{S_j}}+R_C}{R_i+R_C}u_i\right).
	\end{align}
\end{subequations}
The cluster's SoC/OCV relationship is approximated as
\begin{equation}
	\bar{u}_j(\bar{q}_j(t)) =  \bar{\alpha}_j^i(\bar{q}_j(t)) + \bar{\beta}_j^i(\bar{q}_j(t))\bar{q}_j(t),
	\label{ClusterLevelSoCOCV}
\end{equation} 
where 
\begin{equation}
	\bar{\alpha}_j^i = \frac{1}{n_{S_j}}\sum_{j=1}^{n_{S_j}}\alpha_j^i, \quad \bar{\beta}_j^i = 	\frac{1}{n_{S_j}}\sum_{j=1}^{n_{S_j}} \beta_j^i.
\end{equation}
The cluster's internal charging/discharging power can be expressed by
\begin{equation}
	\bar{P}_{S_j} = \bar{u}_j(\bar{q}_j(t))\bar{i}_{L_j}(t).
\end{equation}
Its total power losses can be calculated as
\begin{equation}
	\bar{L}_{S_j} =  \bar{R}_j\bar{i}_{L_j}^2(t).
	\label{ClusterLoss}
\end{equation}
Putting together \eqref{AggregateCapacity}-\eqref{ClusterLoss}, we obtain an electrical model for cluster $S_j$ to grasp the aggregated dynamics of the constituent cells. 
\subsubsection{Thermal Modeling}
This paper uses a lumped thermal model to describe the cells' thermal dynamics \cite{2016-TSTE-PC}. This model assumes that the cell's temperature is concentrated at a single point. The simplification makes it tractable to deal with the cell-to-cluster thermal modeling. Fig.~\ref{CellLevelModel} (b) depicts the model for cell j. It considers the cell's internal power loss, $R_ji_{L_j}^2(t)$, as the source of heat generation and concentrates on the convection between the cell and the environment. The cell-level thermal model can then be expressed by
\begin{equation}
	m_jC_{\textrm{th}}\dot{T}_j(t) = R_ji_{L_j}^2(t) - (T_j(t)-T_{\textrm{env}})/R_{\textrm{conv}},
	\label{CellLevelThermalModel}
\end{equation}
where $T_j$ and $T_{\textrm{env}}$ are the temperature of cell $j$ and the environmental temperature, respectively. Further, $m_j$ and $C_{\textrm{th}}$ are the mass of cell $j$ and specific heat capacity, respectively; $R_{\textrm{conv}}$ is the convective thermal resistance between cell $j$ and the environment, specified by
\begin{equation}
	R_{\textrm{conv}} = \frac{1}{hA_j},
	\nonumber
\end{equation}
where $h$ and $A_j$ are the heat transfer coefficient between the cell's surface and the environment, and the external surface area of cell $j$, respectively.

Given \eqref{CellLevelThermalModel}, we approximate the thermal dynamics of cluster $S_j$ as follows:
\begin{equation}
	\bar{m}_jC_{\textrm{th}}\dot{\bar{T}}_j(t) = \bar{R}_j\bar{i}_{L_j}^2(t) - (\bar{T}_j(t)-T_{\textrm{env}})/\bar{R}_{\textrm{conv},j}.
	\label{ClusterLevelTempDynamic}
\end{equation}
Here, $\bar{T}_j$, $\bar{m}_j$, and $\bar{R}_{\textrm{conv},j}$ are respectively the lumped temperature, mass, and thermal resistance of cluster $S_j$, which are calculated as
\begin{equation}
	\bar{T}_j =\frac{1}{n_{S_j}}\sum_{j=1}^{n_{S_j}}T_j, \ \bar{m}_j = \sum_{j=1}^{n_{S_j}}m_j, \ \bar{R}_{\textrm{conv},j} = \frac{1}{h\sum_{j=1}^{n_{S_j}}A_j}. \nonumber
\end{equation}

From above, we have a cluster-level coupled electro-thermal model. The model has a low-order, compact structure and is computationally amenable to subsequent power optimization.
\subsection{Inter-Cluster Optimal Control}
Based on the cell clustering and cluster modeling, we are now in a good position to deal with the optimal power management for the clusters.
\subsubsection{Problem formulation}
For the considered large-scale BESS, our goal is to minimize the total power losses in operation, while satisfying the physical, safety, and balancing constraints and ensuring the continuous power supply to the load. In pursuit of the goal, we develop an inter-cluster optimization problem. The total power losses for the clusters within the time horizon $[t, t+H)$ are
\begin{equation}
	\int_{t}^{t+H} \bar{L}(\tau)d\tau,
\end{equation}
where $H$ is the horizon length and $\bar{L}(t) = \sum_{j=1}^k \bar{R}_j\bar{i}_{L_j}^2(t)$. We further impose constraints on the charging/discharging currents, temperatures, and SoC of the clusters to guarantee their safe operation as follows:
\begin{subequations}
	\begin{align}
		\bar{i}_{L_j}^{\min} &\leq \bar{i}_{L_j} \leq \bar{i}_{L_j}^{\max}, \label{Safety-1}\\
		\bar{T}_j^{\min} &\leq \bar{T}_j \leq \bar{T}_j^{\max}, \label{Safety-2} \\
		\bar{q}_j^{\min} &\leq \bar{q}_j \leq \bar{q}_j^{\max}, \label{Safety-3}
	\end{align}
	\label{SafetyConstraints}
\end{subequations}
\hspace{-5pt}where $\bar{i}_{L_j}^{\textrm{min}/\textrm{max}}$, $\bar{T}_{_j}^{\textrm{min}/\textrm{max}}$, and $\bar{q}_{_j}^{\textrm{min}/\textrm{max}}$ are the upper/lower safety bounds for the current, temperature, and SoC, respectively. Note that the bounds for $\bar{i}_{L_j}$ depend on $n_{S_j}$, i.e., $\bar{i}_{L_j}^{\textrm{min}/\textrm{max}} = \sum_{i=1}^{n_{S_j}}i_{L_i}^{\textrm{min}/\textrm{max}}$, where $i_{L_i}^{\textrm{min}/\textrm{max}}$ is the upper/lower current bounds for cell $i$. We also enforce cluster-level SoC and temperature balancing constraints as follows:
\begin{subequations}
	\begin{align}
		\left| \bar{q}_j - \bar{q}_{\textrm{avg}} \right| &\leq \Delta \bar{q}, \\
		\left| \bar{T}_j - \bar{T}_{\textrm{avg}} \right| &\leq \Delta \bar{T},
	\end{align}
	\label{BalancingConstraints}
\end{subequations}
\hspace{-5pt}where $\Delta \bar{q}$ and $\Delta \bar{T}$ are the maximum allowed SoC and temperature deviations among the clusters, respectively. The terms $\bar{q}_{\textrm{avg}}$ and $\bar{T}_{\textrm{avg}}$ represent the average SoC and temperature of all the clusters, respectively. Further, the following constraint is introduced to ensure the power supply and demand balance:
\begin{equation}
	\sum_{j=1}^k \bar{P}_{S_j} - \bar{R}_j\bar{i}_{L_j}^2 = P_{\textrm{out}},
	\label{PowerBalanceConstraint}
\end{equation}
where $P_{\textrm{out}}$ is the output power demand. Collecting the cost function and constraints, one can compactly express the optimal power management problem as
\begin{equation}
	\begin{split}
		\min_{\bar{i}_{L_j},j=1,...,k} &\quad \int_{t}^{t+H} \bar{L}(t)dt, \\
		\textrm{s.t.} &\quad \eqref{ClusterLevelSoCDynamic}, \eqref{ClusterLevelTempDynamic}, \eqref{SafetyConstraints}\mbox{-}\eqref{PowerBalanceConstraint}.
	\end{split}
	\label{InterClusterOptimNonConvex}
\end{equation}
This problem seeks to find out the best $\bar{i}_{L_j}$ for cluster $j$ for $j=1,...k$ in a predictive manner over a receding horizon. However, it is a non-convex optimization problem due to the nonlinearity of the equality constraint in \eqref{PowerBalanceConstraint}, thus resisting the search for the global optimum. We adopt and modify the convexification technique in \cite{2012-IFAC-NM} to overcome this issue.

\subsubsection{Convexification of \eqref{InterClusterOptimNonConvex}}

To start with, we define
\begin{equation}
	\bar{E}_j(t) = \frac{1}{2}\bar{C}_j\bar{u}_j^2(\bar{q}_j(t)) - \bar{E}_j^0,
	\label{ClusterLevelEnergy}
\end{equation}
where $\bar{E}_j(t)$ is the remaining energy of cluster $S_j$, $\bar{C}_j=\bar{Q}_j/\bar{\beta_j}$, and $\bar{E}_j^0 = \frac{1}{2}\bar{C}_j\bar{u}_j^2(\bar{q}_j(0))$ is the initial energy. Here, $\bar{E}_j$ is introduced to replace SoC for the purpose of convexification, as will be seen later. Given \eqref{ClusterLevelSoCDynamic}, \eqref{ClusterLevelSoCOCV} and \eqref{ClusterLevelEnergy}, the evolution of $\bar{E}_j$ is governed by
\begin{equation}
	\dot{\bar{E}}_j(t) = -\bar{P}_{S_j}.
\end{equation}
The power losses of cluster $S_j$ can also be expressed in terms of $\bar{E}_j$ and $\bar{P}_{S_j}$ as
\begin{equation}
	\bar{P}_{l_j} = \frac{\bar{R}_j\bar{C}_j\bar{P}_{S_j}^2}{2(\bar{E}_j + \bar{E}_j^0)}.
	\label{ClusterLevelPowerLoss}
\end{equation}
In the convex formulation of the optimal power management problem, we control $\bar{P}_{S_j}$ to minimize the total power losses of the clusters, so \eqref{ClusterLevelPowerLoss} serves as a nonlinear equality constraint resulting in a non-convex problem. Since the power loss term appears in the cost function of the problem, the following relaxation can be considered:
\begin{equation}
	\bar{P}_{l_j} \geq \frac{\bar{R}_j\bar{C}_j\bar{P}_{S_j}^2}{2(\bar{E}_j + \bar{E}_j^0)},
	\label{ClusterLevelPowerLossRelaxed}
\end{equation}
where the optimization problem will reduce $\bar{P}_{l_j}$ to its lower bound. 

Proceeding forward, we can reformulate the safety constraints \eqref{Safety-1} and \eqref{Safety-3} in terms of $\bar{E}_j$ and $\bar{P}_{S_j}$ as follows:
\begin{subequations}
	\begin{align}
		\sqrt{\frac{2}{\bar{C}_j}(\bar{E}_j+\bar{E}_j^0)}\bar{i}_{L_j}^\textrm{min} &\leq \bar{P}_{S_j} \leq \sqrt{\frac{2}{\bar{C}_j}(\bar{E}_j+\bar{E}_j^0)}\bar{i}_{L_j}^\textrm{max}, \\
		\frac{1}{2}\bar{C}_j\bar{u}_j^2(\bar{q}_j^\textrm{min}(t)) &\leq \bar{E}_j+\bar{E}_j^0 \leq \frac{1}{2}\bar{C}_j\bar{u}_j^2(\bar{q}_j^\textrm{max}(t)).
	\end{align}
\end{subequations}
We also rewrite and modify the constraints in \eqref{BalancingConstraints} as 
\begin{subequations}
	\begin{align}
		\Bigg|\frac{2}{\bar{C}_j}\bar{E}_j(t)-\frac{1}{k}\sum_{l=1}^{k}\frac{2}{\bar{C}_l}\bar{E}_l(t)\Bigg|& \leq \Delta \bar{E}_j+\bar{\xi}^{(E)}_j,\\
		\left|\bar{T}_j(t)-\bar{T}_{\textrm{avg}}(t)\right|& \leq \Delta \bar{T}+\bar{\xi}^{(T)}_j,
	\end{align}
	\label{ConvexBalancingConstraints}
\end{subequations}
\hspace{-5pt}where $\Delta \bar{E}_j = (\bar{\alpha}_j+\bar{\beta}_j\Delta \bar{q})^2-\bar{\alpha}_j^2$, and $\bar{\xi}^{(E)}_j$ and $\bar{\xi}^{(T)}_j$ are respectively the energy and temperature slack variables. The addition of the slack variables is to help fix the potential infeasibility issue, which would happen when the clusters face too large variations in their initial SoC and temperature \cite{TTE-FA-2022}. The slack variables will also be included in the cost function to penalize constraint violations. 

Now, we are ready to introduce a convex inter-cluster power optimization problem and present it in a discrete-time form for the sake of computation and implementation. We denote by $\bar{z}_j = \left[\begin{smallmatrix}\bar{P}_{S_j}&\bar{P}_{l_j}&\bar{E}_j&\bar{T}_j&\bar{\xi}_j^{(E)}&\bar{\xi}_j^{(T)}\end{smallmatrix}\right]^\top$ the vector of the optimization variables for $j=1,...,k$. We state the new problem as
\begin{equation}
	\begin{aligned}
		&\min_{\bar{z}_j, j=1,...,k} \quad \sum_{t=0}^{H} \sum_{j=1}^{k} \bar{P}_{l_j}[t] +\bar{\lambda}^{(E)} \bar{\xi}^{(E)}_j[t] + \bar{\lambda}^{(T)} \bar{\xi}^{(T)}_j[t],\\
		&\textrm{Safety constraints:} \quad \\
		&\ \sqrt{\frac{2}{\bar{C}_j}(\bar{E}_j[t]+\bar{E}_j^0)}\bar{i}_{L_j}^\textrm{min} \leq \bar{P}_{S_j}[t] \leq \sqrt{\frac{2}{\bar{C}_j}(\bar{E}_j[t]+\bar{E}_j^0)}\bar{i}_{L_j}^\textrm{max},\\
		&\bar{T}_j^{\textrm{min}}\leq \bar{T}_j\leq \bar{T}_j^{\textrm{max}}\\
		&\ \frac{1}{2}\bar{C}_j\bar{u}_j^2(\bar{q}_j^\textrm{min}[t]) \leq \bar{E}_j[t]+\bar{E}_j^0 \leq \frac{1}{2}\bar{C}_j\bar{u}_j^2(\bar{q}_j^\textrm{max}[t]),\\
		&\textrm{Balancing constraints:} \quad \\
		&\ \Bigg|\frac{2}{\bar{C}_j}\bar{E}_j[t]-\frac{1}{k}\sum_{l=1}^{t}\frac{2}{\bar{C}_l}\bar{E}_l[t]\Bigg| \leq \Delta \bar{E}_j+\bar{\xi}^{(E)}_j[t],\\
		&\ \left|\bar{T}_j[t]-\bar{T}_{\textrm{avg}}[t]\right| \leq \Delta \bar{T}+\bar{\xi}^{(T)}_j[t],\\
		&\textrm{Power loss constraint:} \quad \\
		&\ \bar{P}_{l_j}[t] \geq \frac{\bar{R}_j\bar{C}_j\bar{P}_{S_j}^2[t]}{2(\bar{E}_j[t]+\bar{E}_j^0)},\\
		&\textrm{Energy dynamics:} \quad \\
		&\ \bar{E}_j[t+1]-\bar{E}_j[t]=-\bar{P}_{S_j}[t]\Delta t, \\
		&\textrm{Thermal dynamics:} \quad \\
		&\ \bar{T}_j[t+1]=\bar{T}_j[t] + \frac{\Delta t}{\bar{m}_jC_{\textrm{th}}}\Big[\bar{P}_{l_j}[t]-(\bar{T}_j[t]-T_{\textrm{env}})/\bar{R}_{\textrm{conv}}\Big],\\
		&\textrm{Power supply-demand balance:} \quad \\
		&\ \sum_{j=1}^k \bar{P}_{S_j}[t]-\bar{P}_{l_j}[t]=P_{\textrm{out}}[t],
		\label{InterClusterOptimConvex}
	\end{aligned}
\end{equation}
where $\bar{\lambda}^{(E)}$ and $\bar{\lambda}^{(T)}$ are the penalization weights for $\bar{\xi}^{(E)}$ and $\bar{\xi}^{(T)}$, respectively. The optimization problem in \eqref{InterClusterOptimConvex} is convex because the cost function and the domain are both convex. The convexity would allow efficient computation of the global optimum with well-known algorithms.
\begin{remark}
Cell-based optimal power management for BESS  has been investigated in different studies, including our prior work~\cite{TTE-FA-2022}, but the existing approach face a significant computational bottleneck. For example, one must solve a receding-horizon constrained optimization problem in~\cite{TTE-FA-2022} that is akin to~\eqref{InterClusterOptimConvex} but involves all cells. Its computation will quickly reach a formidable level when the number of cells grows. Our cluster-based approach, as shown in~\eqref{InterClusterOptimConvex}, will be computationally much cheaper. This is because it involves only $6Hk - k$ optimization variables and  $11Hk+5k+H$ constraints with $k \ll n$, contrasting with $6Hn - n$ variables and $11Hn+5n+H$ constraints in~\cite{TTE-FA-2022}, as summarized in Table~\ref{Table_NEW}.  While the clustering and power splitting will add some computation, there is  little compromise to the overall efficiency and scalability of our approach. 
\end{remark}

\begin{table}[!t]
	\renewcommand{\arraystretch}{1.3}
	\caption{Comparison of the Computational Complexity of \eqref{InterClusterOptimConvex}}
	\centering
	\label{Table_NEW}
	\resizebox{\columnwidth}{!}{
		\begin{tabular}{c c c }
			\hline\hline \\[-3mm]
			\multicolumn{1}{c}{Method} & \multicolumn{1}{c}{\# of optimization variables} & \multicolumn{1}{c}{\# of constraints}  \\[1.6ex] \hline
			\cite{TTE-FA-2022} 					& $6Hn-n$ 	&	$11Hn+5n+H$ \\
			\eqref{InterClusterOptimConvex} in our approach 					& $6Hk-k$ 	&	$11Hk+5k+H$ \\	\hline\hline
\multicolumn{1}{l}{$k \ll n$}
		\end{tabular}
	}
\end{table}

We emphasize that, while \eqref{InterClusterOptimConvex} attempts to balance the clusters in SoC and temperature, it is of our interest to make the cells balanced as well. But some cells might be slightly outside the desired balancing bounds due to the aggregation in the cluster modeling. We address this issue by an adaptive choice of $\Delta \bar{E}_j$ and $\Delta \bar{T}$, as discussed next.

\subsubsection{Adaptive cell balancing}

The adaptive selection of balancing bounds has been examined within the framework of two conditions: bound tightening and bound relaxing.

\begin{figure}[!t]
    \centering
    \subfloat[\centering ]{{\includegraphics[width=8cm]{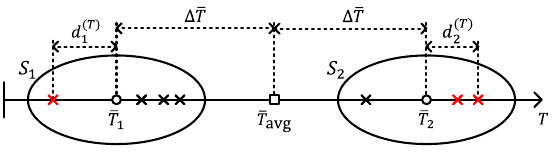} }}
    \;
    \subfloat[\centering ]{{\includegraphics[width=8cm]{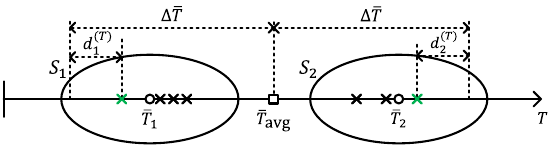} }}
    \caption{Adaptive cell balancing illustration. (a) Bound tightening. (b) Bound relaxing.}
    \label{AdaptiveBound}
\end{figure}

\begin{itemize}
	\item \textit{Bound tightening:} To clarify why a cluster-level balancing constraint may not effectively enforce cell-level balancing, Fig.~\ref{AdaptiveBound} (a) illustrates an example in which seven cells are grouped into two clusters, $S_1$ and $S_2$. The aggregated temperatures of the clusters $S_1$ and $S_2$ are $\bar{T}_1$ and $\bar{T}_2$, respectively. The average of them is $\bar{T}_{\textrm{avg}}$. Suppose that $\left|\bar{T}_{i}-\bar{T}_{\textrm{avg}}\right| = \Delta \bar{T}$ for $i=1,2$ as shown in Fig.~\ref{AdaptiveBound}. Even though the clusters are balanced in this case, some cells (marked in red) are still outside of the desired balancing bounds. As minor as the deviation might be, it is still preferable to maximize the cell-level balance even in the cluster-level optimization procedure. The same phenomenon may also appear in SoC balancing. To address this issue, we suggest to dynamically modify $\Delta \bar{q}$ and $\Delta \bar{T}$ as follows:
\begin{subequations}
	\begin{align}
		\Delta \bar{q}' &= \Delta \bar{q} - \frac{1}{2}\max \{d_1^{(q)},...,d_k^{(q)}\},\\
		\Delta \bar{T}' &= \Delta \bar{T} - \frac{1}{2}\max \{d_1^{(T)},...,d_k^{(T)}\},
	\end{align}
	\label{BoundUpdate}
\end{subequations}
\hspace{-5pt}where $\Delta \bar{q}'$ and $\Delta \bar{T}'$ are the modified cluster-level balancing bounds to replace $\Delta \bar{q}$ and $\Delta \bar{T}$ in \eqref{InterClusterOptimConvex}, and $d_j^{(q/T)}$ is the maximum deviation of a cell's SoC/temperature from the centroid of cluster $S_j$. With \eqref{BoundUpdate}, we practically tighten the balancing bounds for the clusters so as to improve the cell balancing.
	\item \textit{Bound relaxing:} This condition involves adjusting the bounds to expand the permissible range of cluster-level balancing. As illustrated in Figure~\ref{AdaptiveBound} (b), the temperature distribution of cells within clusters $S_1$ and $S_2$ adheres to the desired bound of $\Delta \bar{T}$. To enhance power loss minimization, the balancing bounds for the clusters are relaxed, providing the optimization process with increased flexibility. The relaxation of bounds is determined by utilizing \eqref{BoundUpdate}, where $d_j^{(q,T)} \leq 0$.
\end{itemize}

Note that we do not update the balancing bounds at every time instant. Instead, \eqref{BoundUpdate} is applied only when $\bar{\xi}_{j}^{(E)}$ and $\bar{\xi}_{j}^{(T)}$ reach zero---at this moment, the clusters are balanced so that it is tenable to narrow/relax the bounds. The adaptive cell balancing procedure will be examined in the simulation results.
\subsection{Cluster-to-Cell Power Split}
Suppose that the optimal power quota $\bar{P}_{S_j}^*$ is obtained for cluster $j$ for $j=1,...k$ by solving the optimization problem in \eqref{InterClusterOptimConvex}. The remaining question is how to split $\bar{P}_{S_j}^*$ among the cells within the cluster. We propose three schemes that derive from different assumptions about the intra-cluster cell discrepancy.

\textit{Scheme \#1: Equal power split.} The first scheme assumes that the constituent cells of the clusters are approximately identical, neglecting discrepancies among them in terms of SoC, temperature and internal resistance. It is thus sensible to equally divide the optimal power assigned to a cluster among the constituent cells. For cluster $S_j$, we have
\begin{equation}
	P_{b_1}=...=P_{b_{n_{S_j}}}=\frac{\bar{P}_{S_j}^*}{n_{S_j}}.
\end{equation}

\textit{Scheme \#2: Internal-resistance-based power split.} The second scheme considers the internal resistance variations among the cells and aims to reduce the power losses by an uneven distribution of the cluster power. It divides $\bar{P}_{S_j}^*$ by
\begin{equation}
	P_{b_i} = \frac{\frac{1}{R_i}\bar{P}_{S_j}^*}{\frac{1}{R_1}+\frac{1}{R_2}+...+\frac{1}{R_{n_{S_j}}}}, \quad i=1,...,n_{S_j}.	
\end{equation} 
From above, the less $R_i$ is, the larger $P_{b_i}$ is. 

Compared to scheme \#1, this scheme prioritizes loss minimization over SoC and temperature balancing. However, both of them are based on heuristics, implying suboptimal cell-level power assignment. One can perform intra-cluster optimization to achieve optimal power split among the cells.

\textit{Scheme \#3: Optimal power split.} We can apply the idea in \eqref{InterClusterOptimConvex} to achieve optimal power split by minimizing the power losses of all the cells within the cluster under some safety and SoC balancing constraints. Considering cluster $S_j$, the total power losses can be expressed as
\begin{equation}
	L_j(t) = \sum_{l=1}^{n_{S_j}} (R_l+R_C)i_{L_l}^2(t).
\end{equation}
We enforce the following constraints to ensure safety and intra-cluster balancing:
\begin{subequations}
	\begin{align}
		i_{L_l}^{\min} \leq i_{L_l} &\leq i_{L_l}^{\max}, \label{IntraClusterSafety-1}\\
		T_l^{\min} \leq T_l &\leq T_l^{\max}, \label{IntraClusterSafety-2} \\
		q_l^{\min} \leq q_l &\leq q_l^{\max}, \label{IntraClusterSafety-3} \\
		\left| q_l - q_{j,\textrm{avg}} \right| &\leq \Delta q, \\
		\left| T_l - T_{j,\textrm{avg}} \right| &\leq \Delta T,
	\end{align}
	\label{IntraClusterSafety-BalancingConstraints}
\end{subequations}
\hspace{-5pt}for $l=1,...,n_{S_j}$, where $T_{l}^{\textrm{min}/\textrm{max}} = \bar{T}_{j}^{\textrm{min}/\textrm{max}}$ and $q_{l}^{\textrm{min}/\textrm{max}} = \bar{q}_{j}^{\textrm{min}/\textrm{max}}$, and $q_{j,\textrm{avg}}$ and $T_{j,\textrm{avg}}$ are the average SoC and temperature for cluster $j$, respectively. Further, we must make the total power of the cells equal to the cluster's optimal power. That is, 
\begin{equation}
	\sum_{l=1}^{n_{S_j}} \left(P_{b_l}-(R_l+R_C)i_{L_l}^2\right)=\bar{P}_{S_j}^* - \bar{P}_{l_j}^*.
	\label{IntraClusterPowerBalanceConstraint}
\end{equation}
We can then derive the intra-cluster optimization problem as:
\begin{equation}
	\begin{split}
		\min_{i_{L_l},l=1,...,n_{S_j}} &\quad L_j(t), \\
		\textrm{subject to} &\quad \eqref{ClusterLevelSoCDynamic}, \eqref{ClusterLevelTempDynamic}, \eqref{IntraClusterSafety-BalancingConstraints}, \eqref{IntraClusterPowerBalanceConstraint}.
	\end{split}
	\label{IntraClusterOptimNonConvex}
\end{equation}
For the constituent cells of each cluster, the above nonlinear continuous optimization problem aims to find out the best currents $i_{L_l}$ for $l=1,\dots,n_{S_j}$. Note that \eqref{IntraClusterOptimNonConvex}, similar to \eqref{InterClusterOptimNonConvex}, is non-convex due to the nonlinearity of the equality constraint \eqref{IntraClusterPowerBalanceConstraint}. We can follow the same convexification procedure as in \eqref{InterClusterOptimConvex} to formulate an alternative convex optimization problem to solve \eqref{IntraClusterOptimNonConvex}. Considering cluster $S_j$, the problem is stated as below:
\begin{equation}
	\begin{split}
		&\min_{z_l, l=1,...,n_{S_j}} \quad \sum_{l=1}^{n_{S_j}} P_{l_l}[t] +\lambda^{(E)} \xi^{(E)}_l[t] + \lambda^{(T)} \xi^{(T)}_l[t],\\
		&\textrm{Safety constraints:} \quad \\
		&\ \sqrt{\frac{2}{C_l}(E_l[t]+E_l^0)}i_{L_l}^\textrm{min} \leq P_{b_l}[t] \leq \sqrt{\frac{2}{C_l}(E_l[t]+E_l^0)}i_{L_l}^\textrm{max},\\
		&T_l^{\textrm{min}}\leq T_l\leq T_l^{\textrm{max}}\\
		&\ \frac{1}{2}C_lu_l^2(q_l^\textrm{min}[t]) \leq E_l[k]+E_l^0 \leq \frac{1}{2}C_lu_l^2(q_l^\textrm{max}[t]),\\
		&\textrm{Balancing constraints:} \quad \\
		&\ \Bigg|\frac{2}{C_l}E_l[t]-\frac{1}{n_{S_j}}\sum_{l=1}^{n_{S_j}}\frac{2}{C_l}E_l[t]\Bigg| \leq \Delta E_l+\xi^{(E)}_l[t],\\
		&\ \left|T_l[t]-T_{\textrm{avg}}[t]\right| \leq \Delta T+\xi^{(T)}_l[t],\\
		&\textrm{Power loss constraint:} \quad \\
		&\ P_{l_l}[t] \geq \frac{(R_l+R_C)C_lP_{b_l}^2[t]}{2(E_l[t]+E_l^0)},\\
		&\textrm{Energy dynamics:} \quad \\
		&\ E_l[t+1]-E_l[t]=-P_{b_l}[t]\Delta t, \\
		&\textrm{Thermal dynamics:} \quad \\
		&\ T_l[t+1]=T_l[t] + \frac{\Delta t}{m_lC_{\textrm{th}}}\Big[P_{l_l}[t]-(T_l[t]-T_{\textrm{env}})/R_{\textrm{conv}}\Big],\\
		&\textrm{Power supply-demand balance:} \quad \\
		&\ \sum_{l=1}^{n_{S_j}} \left(P_{b_l}[t]-P_{l_l}[t]\right) = \bar{P}^*_{S_j}[t] - \bar{P}^*_{l_j}[t],
		\label{IntraClusterOptim}
	\end{split}
\end{equation}
where $z_l = \left[\begin{smallmatrix}P_{b_l}&P_{l_l}&E_l&T_l&\xi_l^{(E)}&\xi_l^{(T)}\end{smallmatrix}\right]^\top$ collects the optimization variables, $E_l$ denotes the remaining energy in cell $l$, $\xi_l^{(E)}$ and $\xi_l^{(T)}$ represent the energy and temperature slack variables with associated penalizing weights $\lambda^{(E)}$ and $\lambda^{(T)}$ in the cost function. We highlight that scheme \#3 must address the above intra-cluster optimization problems for each cluster and thus requires more computation. However, each such problem involves only $6n_{S_j}$ optimization variables, and independent of each other, they can be parallelized in computing. Thus, scheme \#3 can still be manageable if there are enough computing resources.

\section{Simulation Results}
In this section, we carry out simulations for a BESS consisting of 400 cells. Table \ref{Table_1} summarizes the specifications of the BESS and the parameters of the proposed optimal power management approach. The battery cells are based on Samsung INR18650-25R whose parameters have been identified and reported in \cite{2020-JES-NT}. The profile of the output power $P_{\textrm{out}}$ is based on the Urban Dynamometer Driving Schedule (UDDS) with the peak charging and discharging power of 6 kW and 10 kW, respectively. The simulations simulate the BESS operation for 2,400 seconds, in which the proposed approach runs with a 10-second predictive horizon. We use the CVX package \cite{CVX-1,CVX-2} with Matlab to solve the convex optimization problems in \eqref{InterClusterOptimConvex} and \eqref{IntraClusterOptim} using a workstation with a 3.5GHz Intel Core i9-10920X CPU and 128GB of RAM.

We initialize the cells by randomly sampling their initial SoC, temperature, and internal resistance from the uniform distributions $\mathcal{U}(0.7, 0.75)$, $\mathcal{U}(301, 305)\, \textrm{K}$, and $\mathcal{U}(31.3, 41.3)\, \textrm{m}\Omega$, respectively. The randomized variations will lead to many possible combinations within the cells at the clustering stage, allowing us to better investigate the efficacy of the proposed approach in handling the cell heterogeneities. Further, we conduct the simulations with schemes \#1-3 for the cluster-to-cell power split to compare their respective performances. 

\begin{table}[!t]
	\renewcommand{\arraystretch}{1.3}
	\caption{Specifications of the considered large-scale BESS}
	\centering
	\label{Table_1}
	\resizebox{\columnwidth}{!}{
		\begin{tabular}{l l l}
			\hline\hline \\[-3mm]
			\multicolumn{1}{c}{Symbol} & \multicolumn{1}{c}{Parameter} & \multicolumn{1}{c}{Value [Unit]}  \\[1.6ex] \hline
			$ n $ & Number of battery cells & 400 \\
			$ v $  & Cell nominal voltage & 3.6     [V] \\
			$ Q $ & Cell nominal capacity & 2.5     [Ah] \\ 
			$ R $ & Cell internal resistance & 31.3     [m$\Omega$] \\
			$ [q^{\textrm{min}},q^{\textrm{max}}] $ & Cell SoC limits  & [0.05,0.95] \\ 
			$ [i^{\textrm{min}},i^{\textrm{max}}] $ & Cell current limits & [-7.5,7.5]     [A] \\ 
			$ C_{\textrm{th}} $ & Specific thermal capacitance & 918.49     [J/(K.Kg)] \\ 
			$ m $ & Cell Mass & 0.0438     [Kg] \\ 
			$ A $ & External surface area & 0.0042     [m\textsuperscript{2}] \\ 
			$ h $ & Convection heat transfer coefficient & 5.8     [W/(K.m\textsuperscript{2})] \\ 
			$ R_{\textrm{conv}} $ & Convection thermal resistance & 41.05     [K/W] \\ 
			$ T_{\textrm{env}} $ & Environment temperature & 298     [K] \\ 
			$ \Delta \bar{q} $ & SoC balancing threshold & 0.5\% \\
			$ \Delta \bar{T} $ & Temperature balancing threshold & 0.5    [K] \\
			$ \Delta t $ & Sampling time & 1     [s]\\
			$ H $ & Horizon length & 10     [s]\\
			\hline\hline
		\end{tabular}
	}
\end{table}

\begin{figure*}[t]
	    \centering
    \subfloat[\centering ]{{\includegraphics[trim={1.8cm 0 3cm 1cm},clip,width=5.5cm]{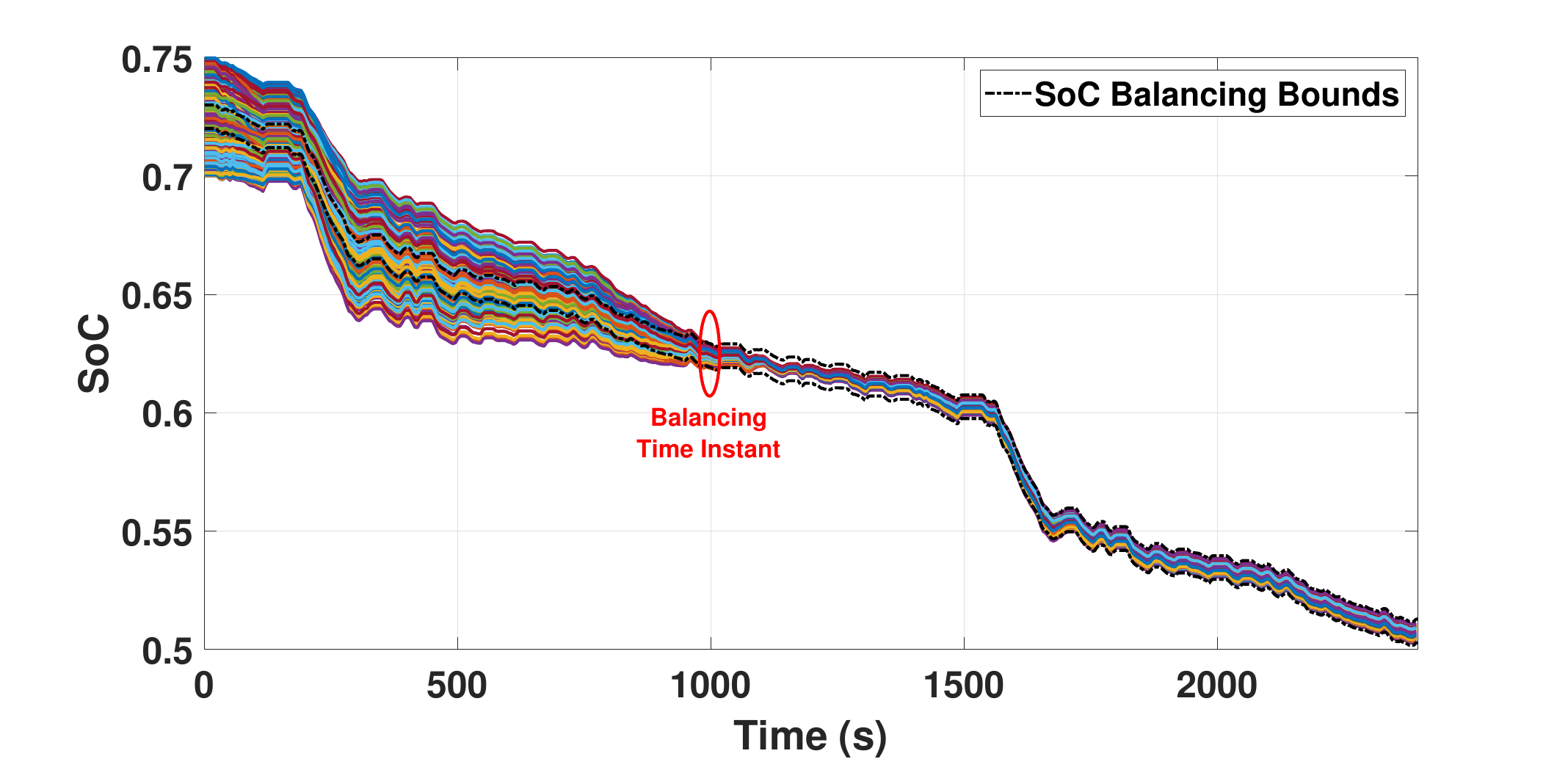} }}
	\,
    \subfloat[\centering ]{{\includegraphics[trim={1.8cm 0 3cm 1cm},clip,width=5.5cm]{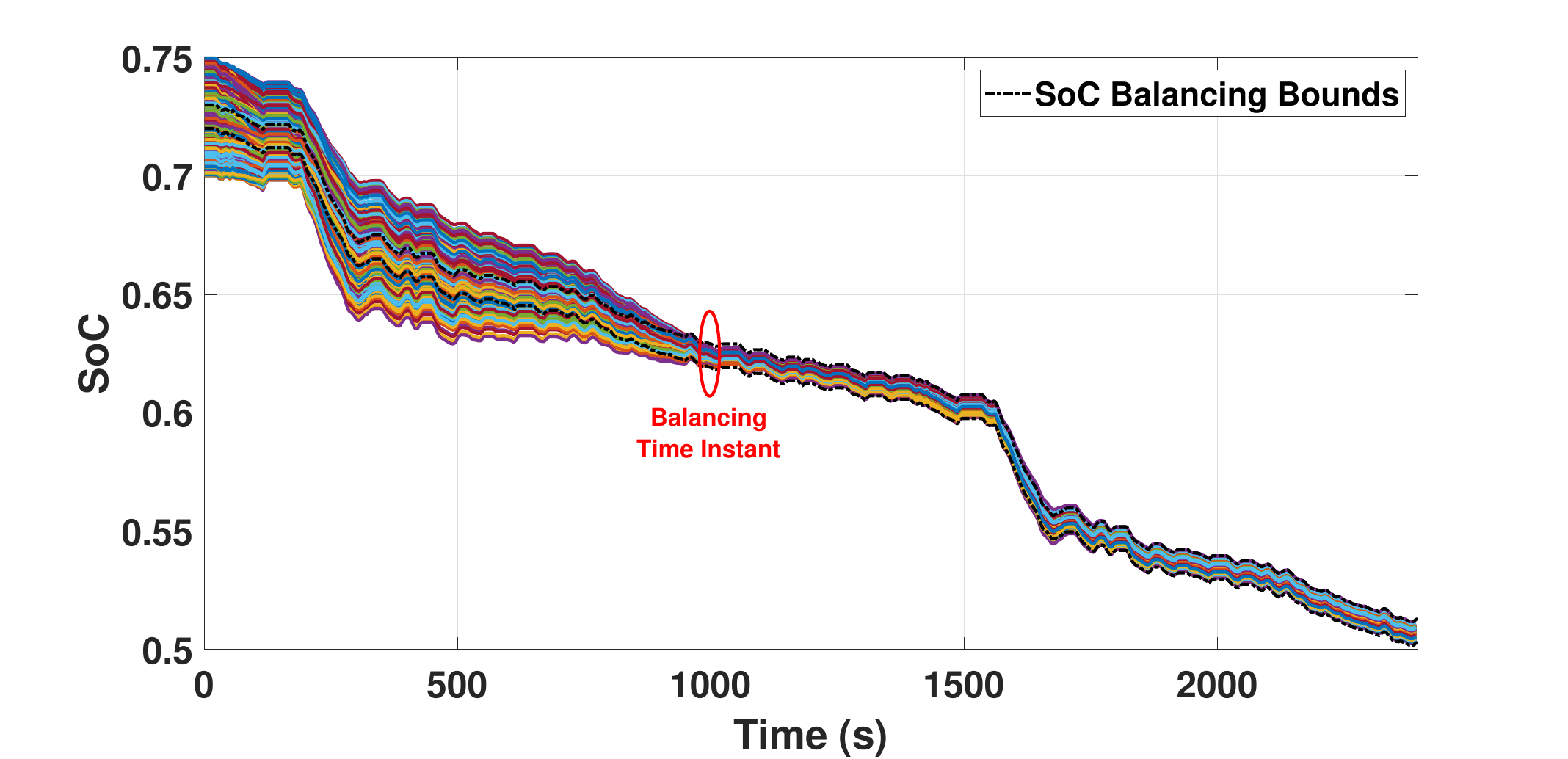} }}
    \,
    \subfloat[\centering ]{{\includegraphics[trim={1.8cm 0 3cm 1cm},clip,width=5.5cm]{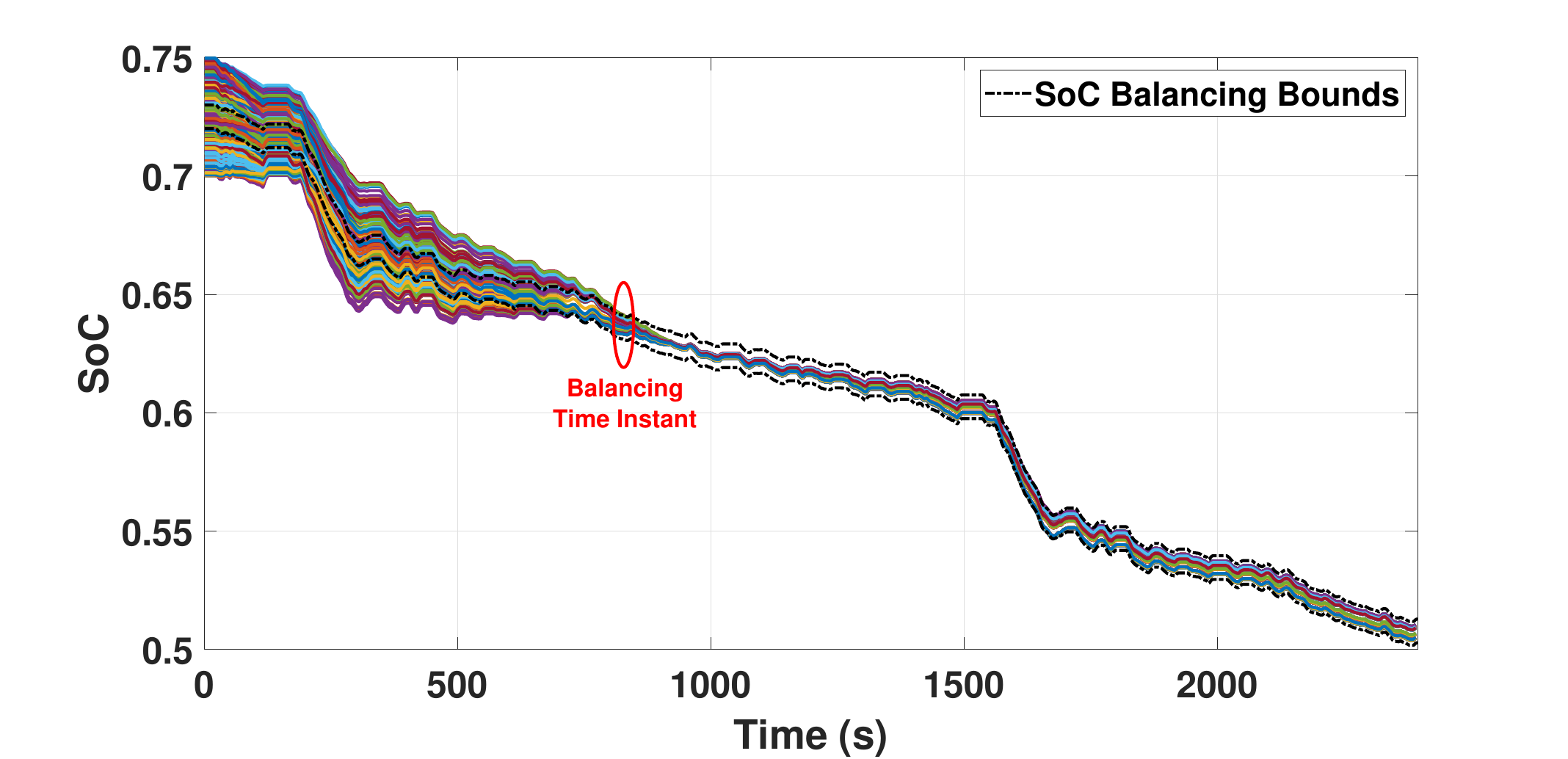} }}
    \,
    \subfloat[\centering ]{{\includegraphics[trim={1.5cm 0 3cm 1cm},clip,width=5.5cm]{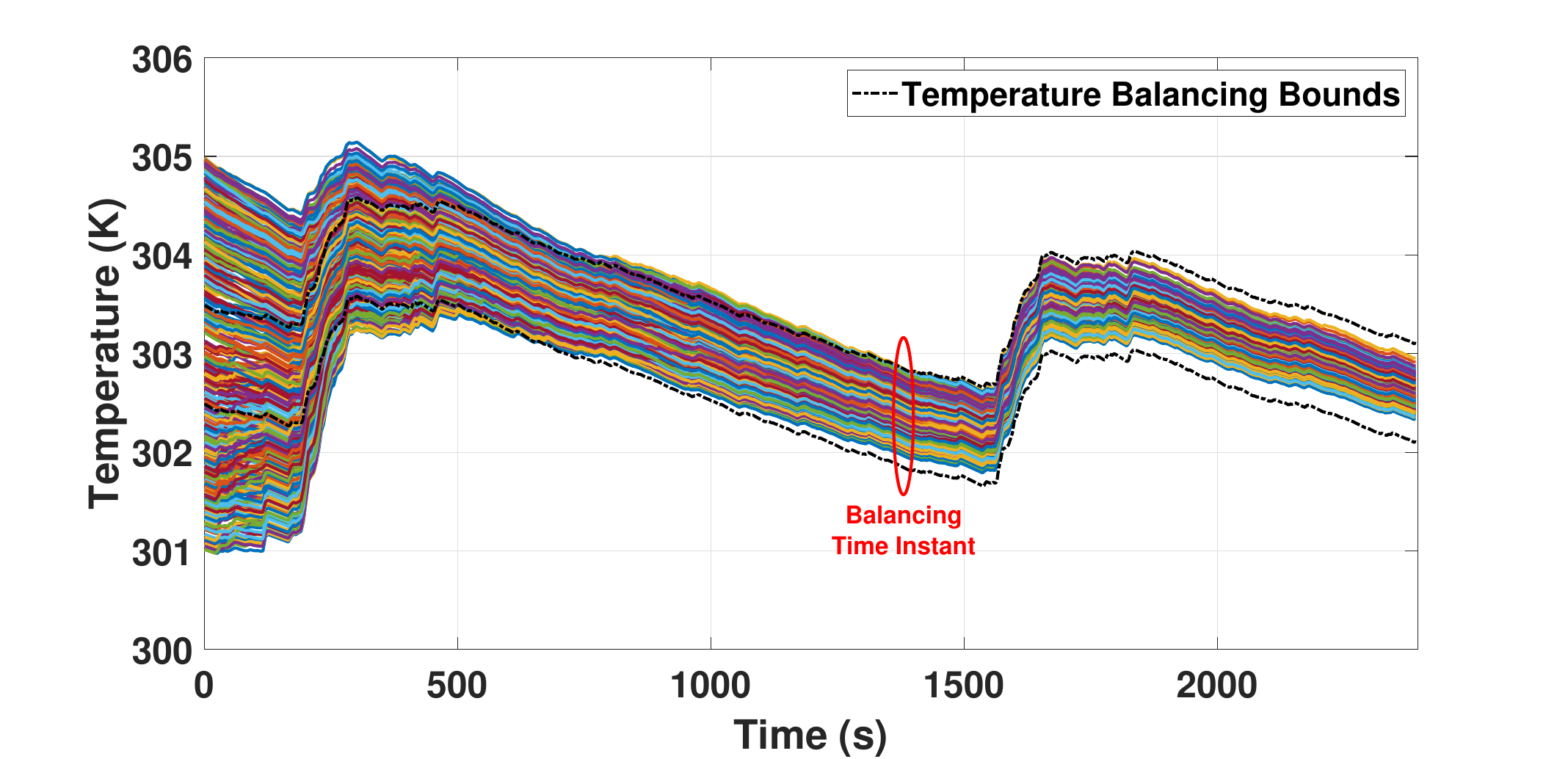} }}
    \,
    \subfloat[\centering ]{{\includegraphics[trim={1.5cm 0 3cm 1cm},clip,width=5.5cm]{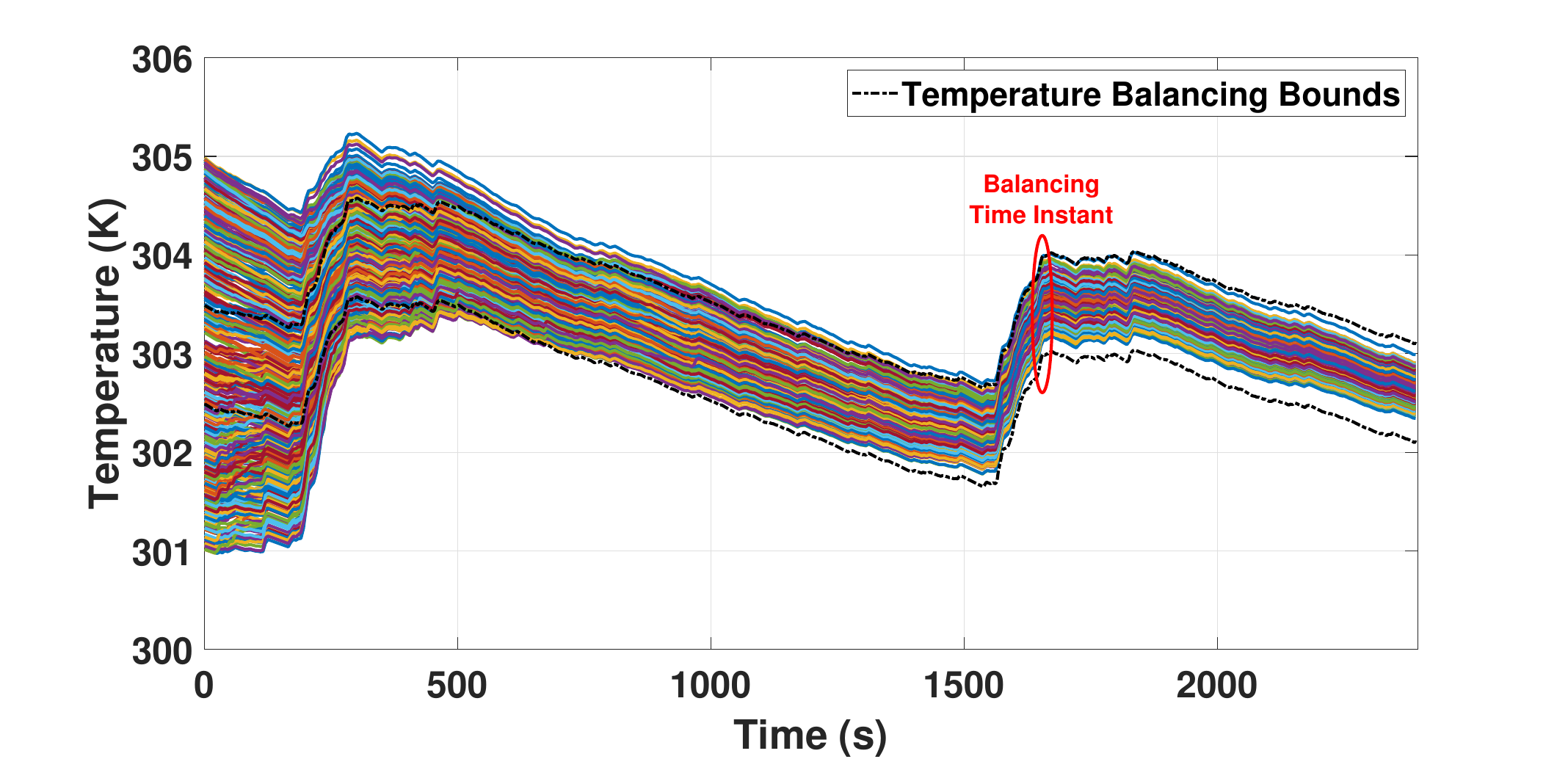} }}
     \,
    \subfloat[\centering ]{{\includegraphics[trim={1.5cm 0 3cm 1cm},clip,width=5.5cm]{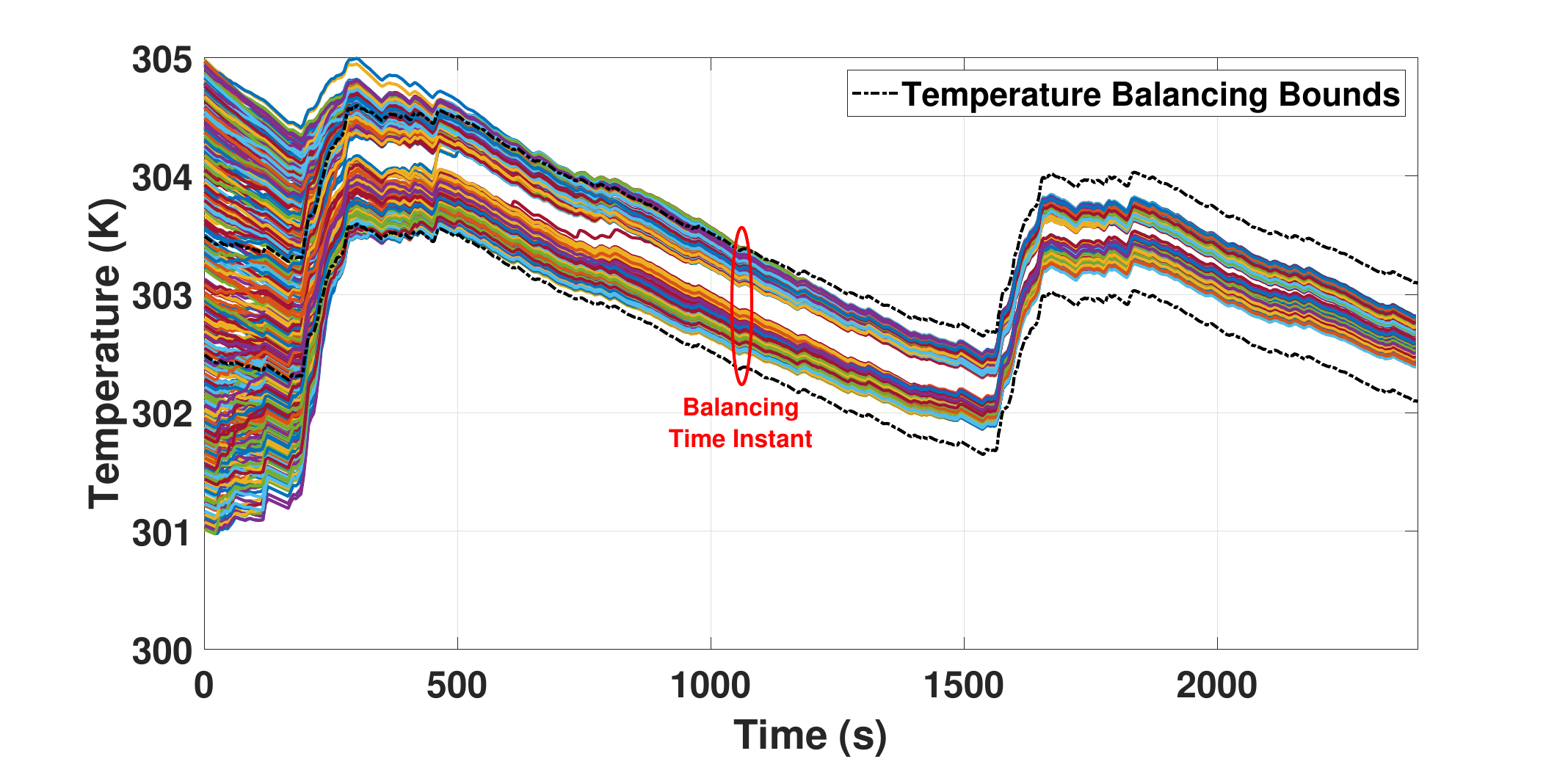} }}
    \caption{Simulation results of the SoC and temperature balancing. (a) The SoC of the cells in scheme \#1. (b) The SoC of the cells in scheme \#2. (c) The SoC of the cells in scheme \#3. (d) The temperature  of the cells in scheme \#1. (e) The temperature  of the cells in scheme \#2. (f) The temperature  of the cells in scheme \#3.}
    \label{FIG_SIM_1}
\end{figure*}

Fig.~\ref{FIG_SIM_1} depicts the SoC and temperature balancing performance of the proposed approach. The proposed approach under three power split schemes effectively alleviates the SoC and temperature unbalance among the cells and successfully drives the cells within the desired bounds. Having a closer look at Figs.~\ref{FIG_SIM_1} (a)-(c), we see that schemes \#1 and \#2 show similar SoC balancing performances, and the proposed approach drives the cells within the desired bound after about 1,000 seconds. Scheme \#3 shows superior performance, and the cell balancing is achieved only after 700 seconds. A similar performance is observed for the temperature balancing in Figs.~\ref{FIG_SIM_1} (d)-(f), where scheme \#3 outperforms schemes \#1 and \#2 with a balancing time of 1,100 seconds versus 1,400 and 1,700 seconds. To conclude, schemes \#1 and \#2 provide acceptable performance and high computational efficiency, and scheme \#3 would deliver the best balancing performance while requiring more computation. A user may choose one of them in practice based on the performance expectations and computing resources.

\begin{figure}[!t]
\centering
\includegraphics[trim={2.8cm 0cm 2.8cm 1cm},clip,width=8.5cm]{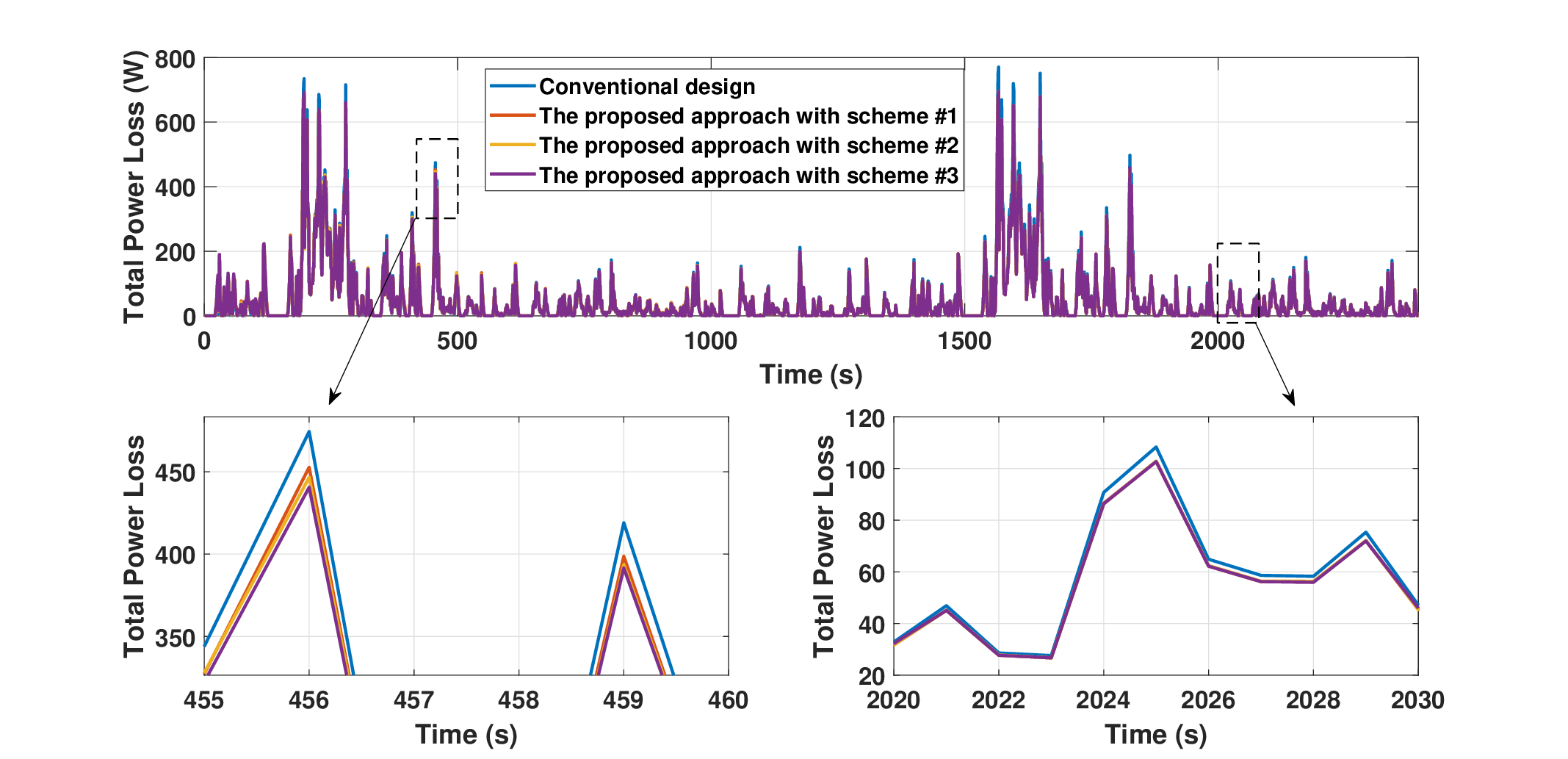}
\caption{The power loss comparison.}
\label{FIG_SIM_2_new_1}
\end{figure}

\begin{figure}[!t]
\centering
\includegraphics[trim={2.5cm 0cm 2.8cm 0cm},clip,width=8.5cm]{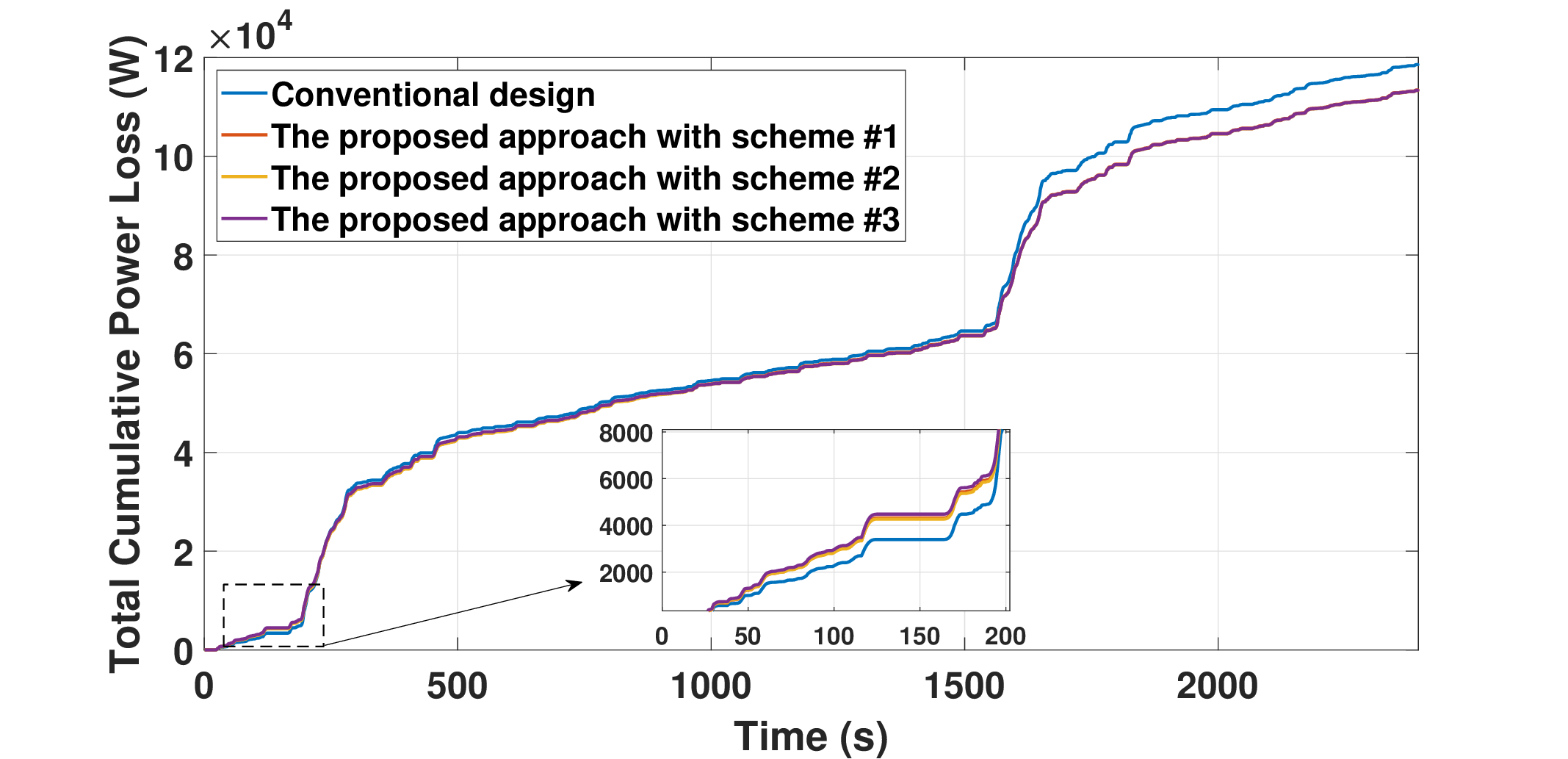}
\caption{The total cumulative power loss.}
\label{FIG_SIM_2_new_2}
\end{figure}

\begin{figure}[!t]
\centering
\includegraphics[trim={2.5cm 0cm 3.2cm 1cm},clip,width=8.5cm]{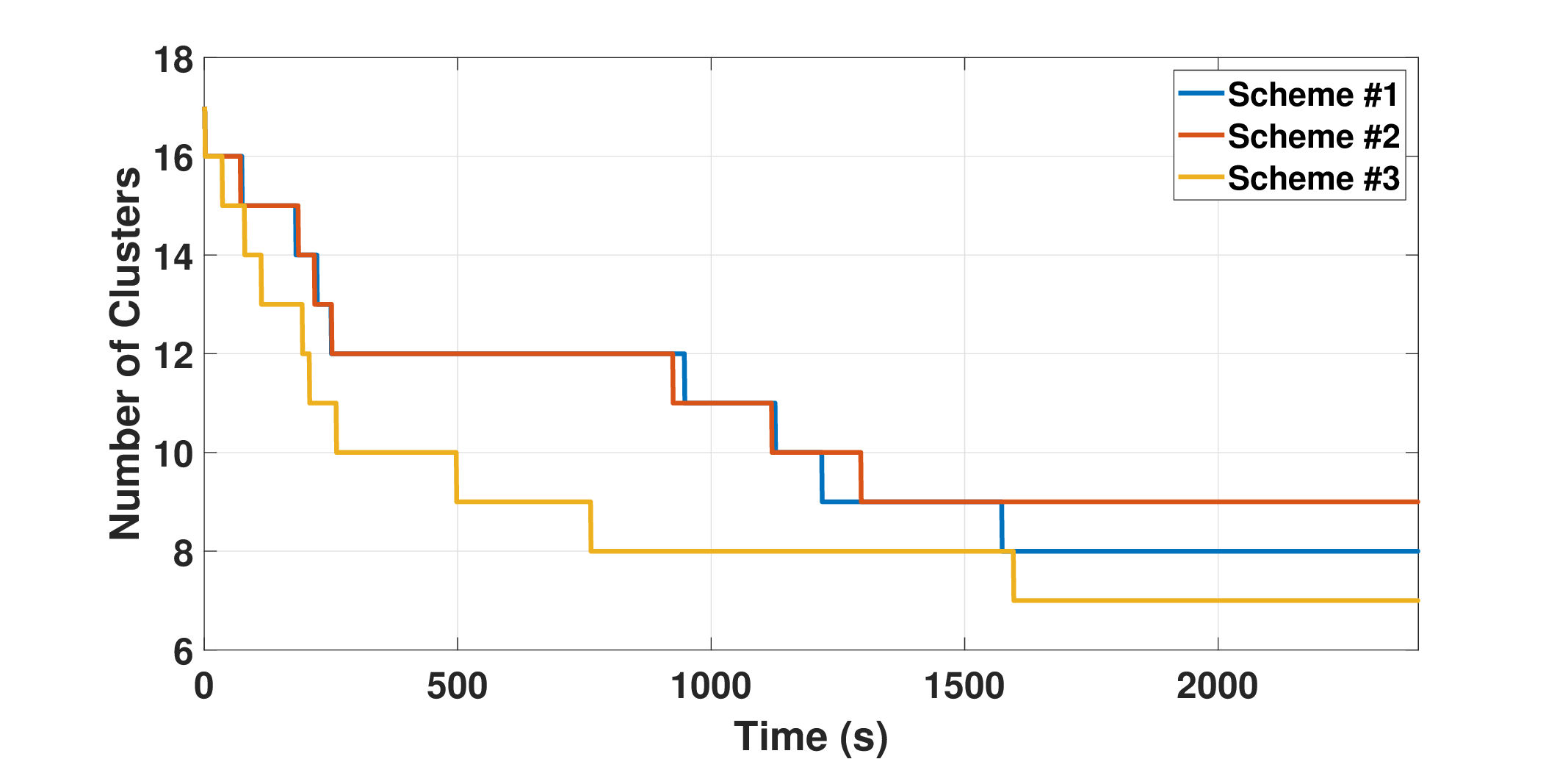}
\caption{Evolution in the number of clusters.}
\label{FIG_SIM_2}
\end{figure}

Fig.~\ref{FIG_SIM_2_new_1} shows the power loss profiles of the BESS under the proposed approach with the three power split schemes, while comparing with the conventional case of no cell-level power control. Overall, we observe the proposed approach leads to lower power losses than in the conventional case. The magnified views in the two time windows of Fig.~\ref{FIG_SIM_2_new_1} show that scheme \#3 causes conspicuously minimum power losses when the cells are unbalanced and that the three schemes produce almost the same power losses when the cells are balanced. Fig.~\ref{FIG_SIM_2_new_2} shows the total cumulative power losses. It is interesting to observe that the proposed approach bears more power losses for the sake of cell balancing in the early time interval of 0-200 seconds. However, its cumulative power loss will become less as time goes by and lower by 4.6\% than in the conventional case at the end of the simulation run. 

Fig.~\ref{FIG_SIM_2} illustrates the change in the number of clusters $k$ throughout the simulation run. The value of $k$ depends on the level of variations in the cells' SoC, temperature, and internal resistance. The more balanced the cells, the fewer the clusters. We can see that the cells are initially grouped into 17 clusters, and fewer clusters are formed as the cells become more uniform. Note that the number of clusters also determines the computational costs of the inter-cluster optimization in \eqref{InterClusterOptimConvex}, so computation becomes less as the simulation runs forward. 

\begin{figure}[!t]
    \centering
    \subfloat[\centering ]{{\includegraphics[trim={2.4cm 0cm 3.2cm 1cm},clip,width=8.5cm]{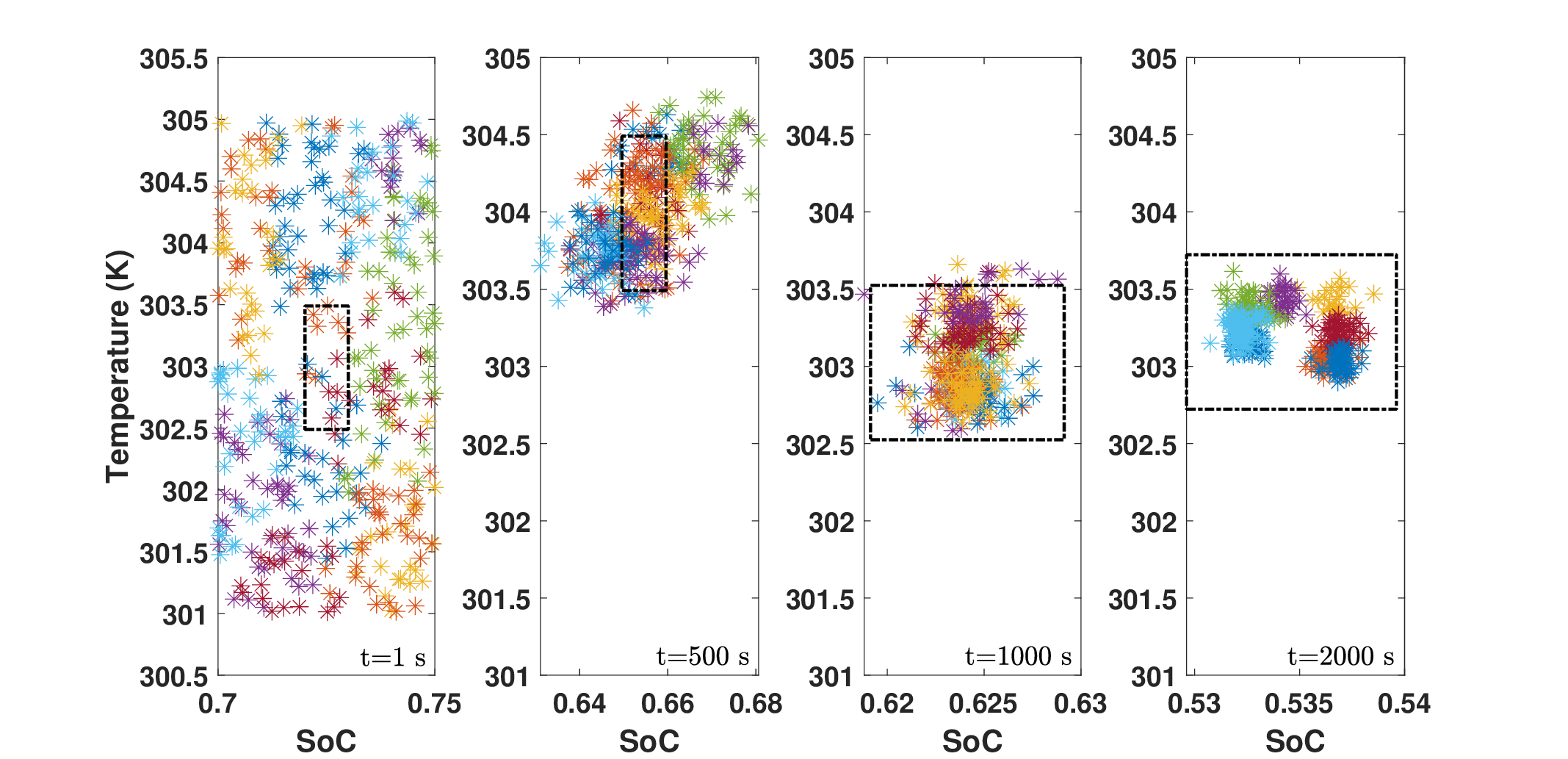} }}
    \;
    \subfloat[\centering ]{{\includegraphics[trim={2.4cm 0cm 3.2cm 1cm},clip,width=8.5cm]{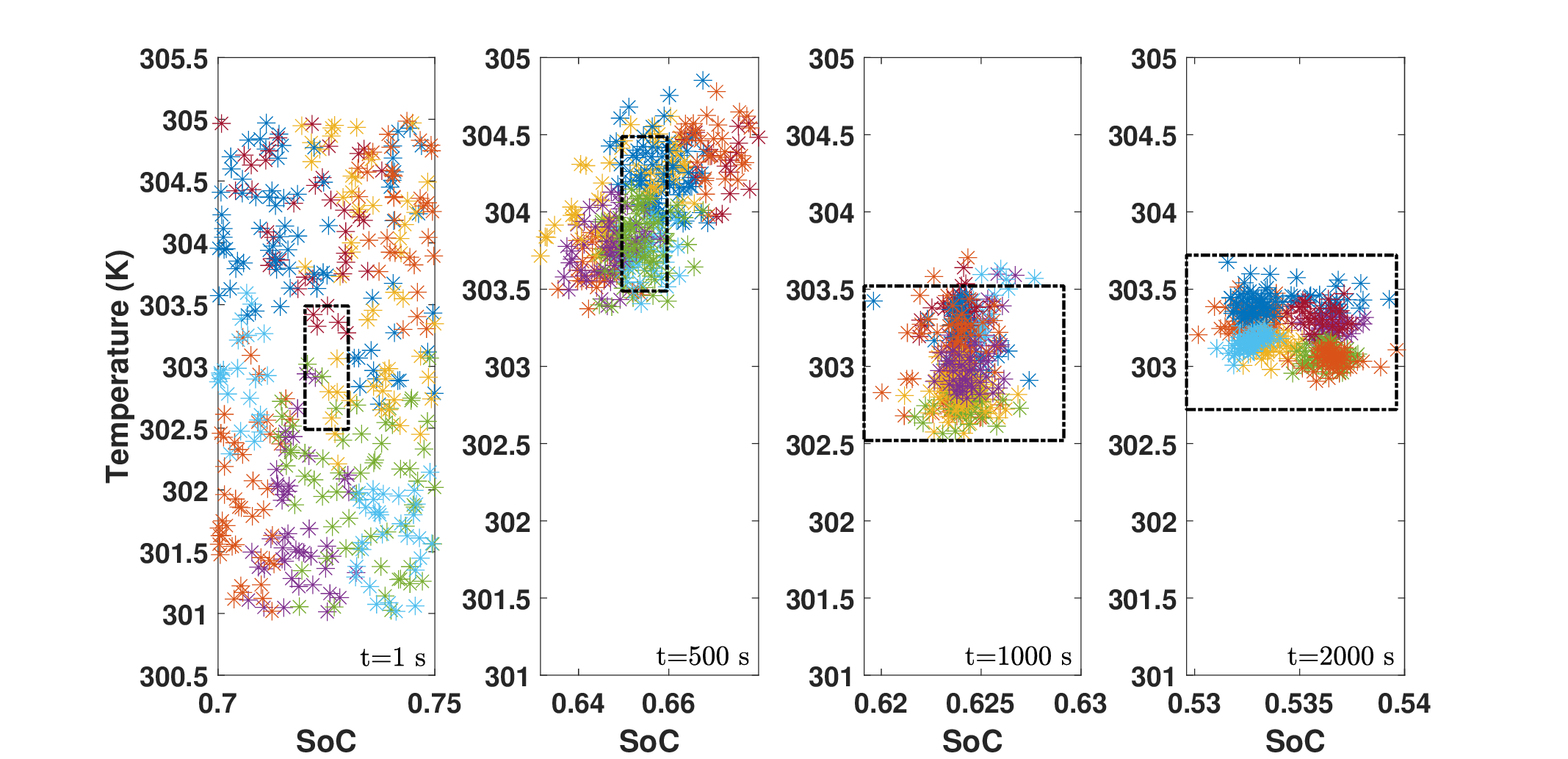} }}
    \;
    \subfloat[\centering ]{{\includegraphics[trim={2.4cm 0cm 3.2cm 1cm},clip,width=8.5cm]{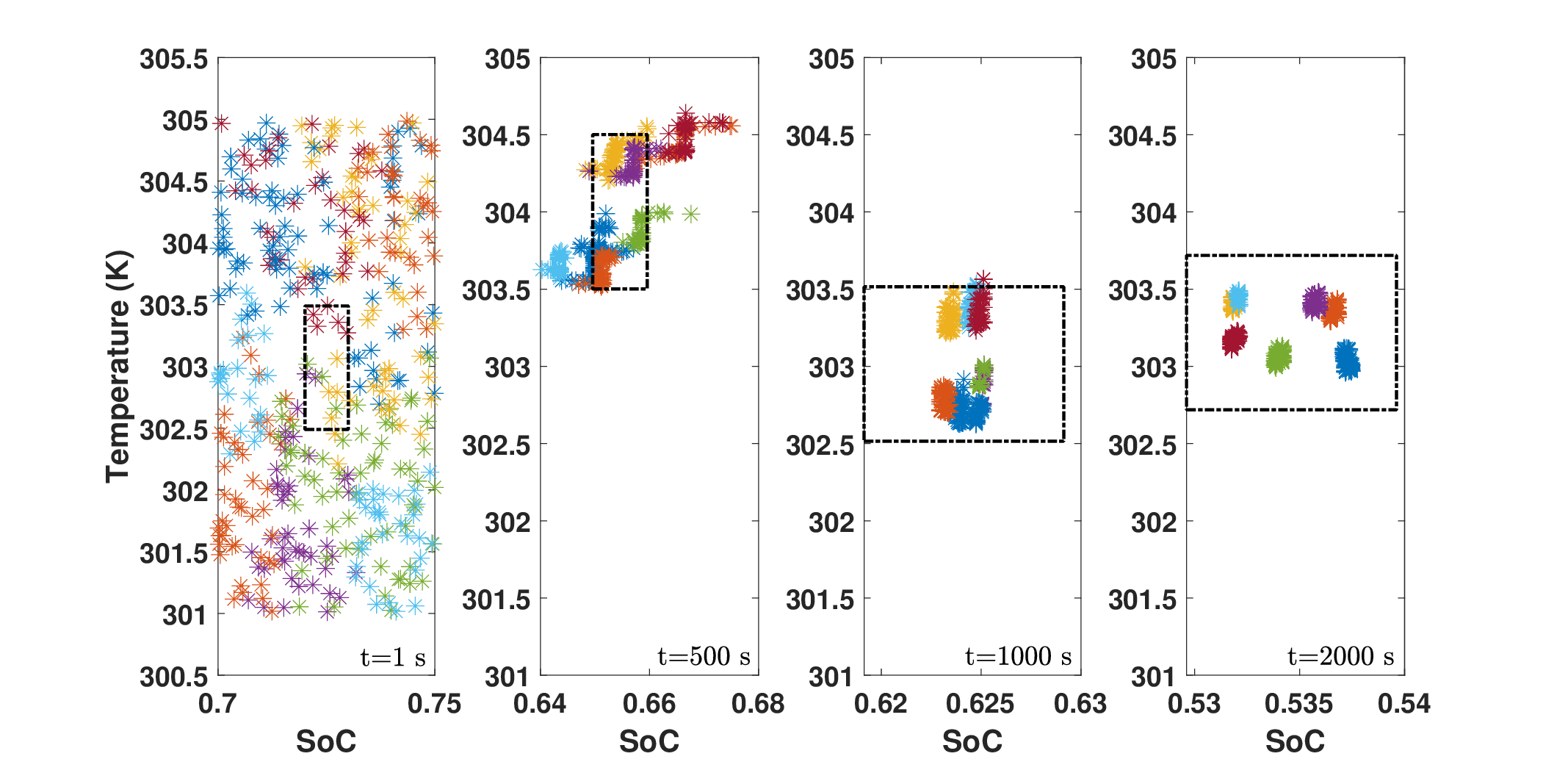} }}
    \caption{The evolution of cells' SoC, temperature, and the clustering procedure. (a) Scheme \#1. (b) Scheme \#2. (c) Scheme \#3.}
    \label{FIG_SIM_3}
\end{figure}

It is of our interest to observe the evolution of the cells' SoC and temperature. Fig.~\ref{FIG_SIM_3} provides snapshots from the cell's SoC and temperature under the three power split schemes at the time instants of 1,500, 1,000, and 2,000 seconds. The black boxes in the plots show the desired SoC and temperature balancing bounds, and each color represents a cluster. The proposed optimal power management approach effectively categorizes the cells into clusters, performs inter-cluster optimization, and drives the cells into the desired bound under all three power split schemes. Taking a closer look at Fig.~\ref{FIG_SIM_3} (c) at the time instant of 2,000 seconds, one can see that the 400 cells form seven distinct clusters. However, the cells are more scattered at the same time instant for schemes \#1 and \#2, while scheme \#3 leads to better intra-cluster convergence. 

\begin{figure}[!t]
    \centering
    \subfloat[\centering ]{{\includegraphics[trim={1.8cm 0cm 3.2cm 1cm},clip,width=8.5cm]{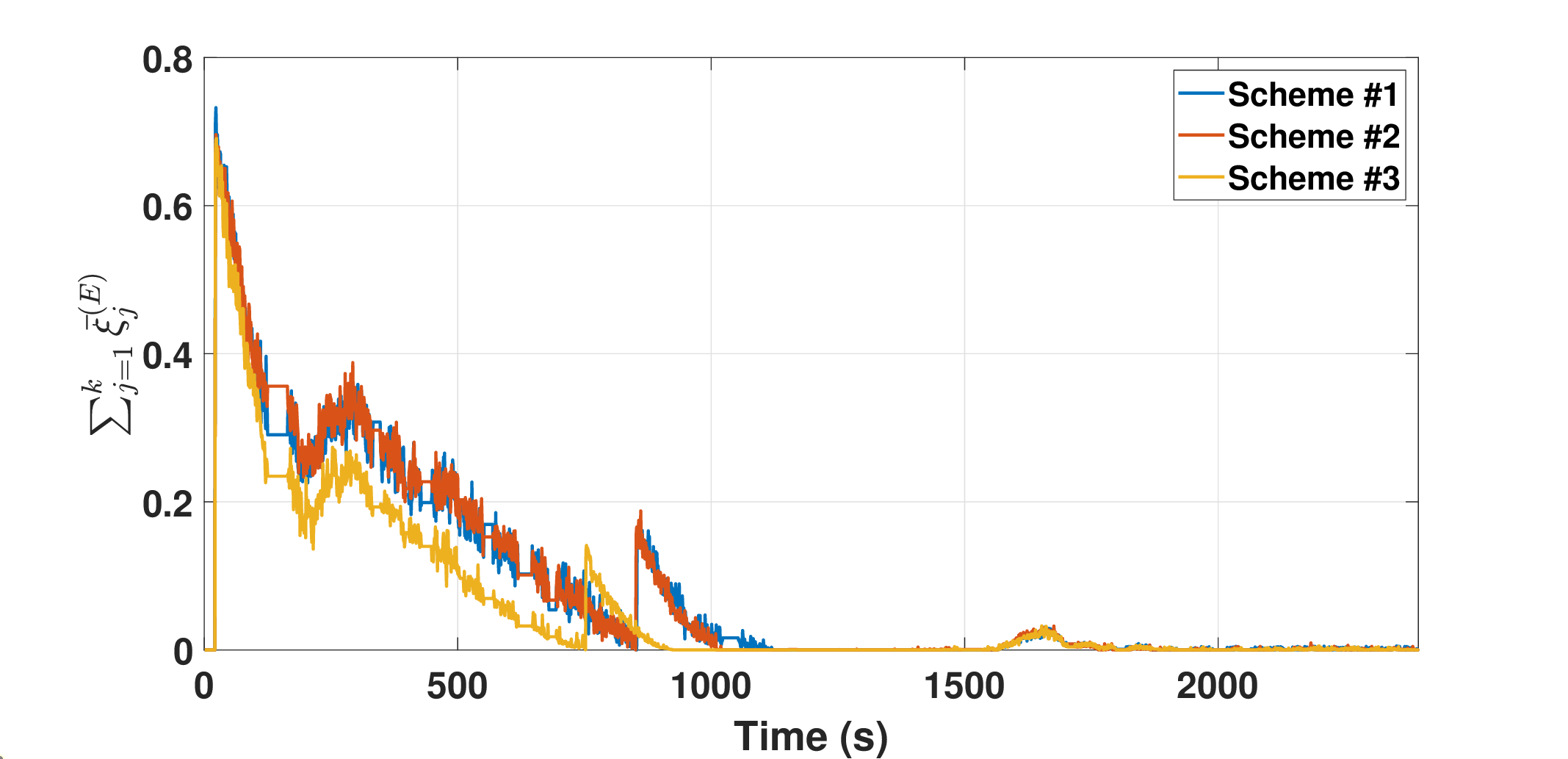} }}
    \;
    \subfloat[\centering ]{{\includegraphics[trim={2.1cm 0cm 3.2cm 1cm},clip,width=8.5cm]{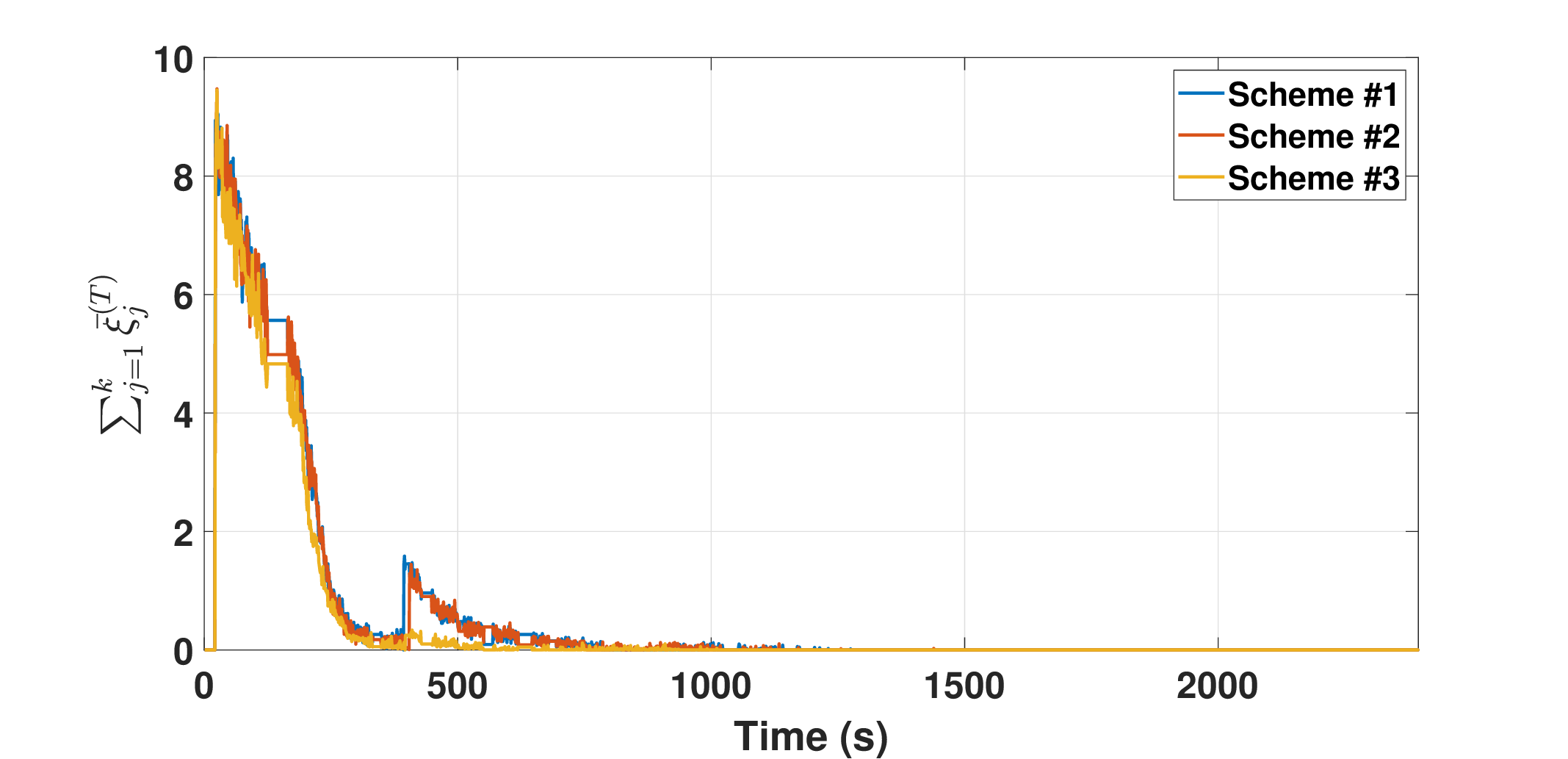} }}
    \;
    \caption{The evolution of the slack variables. (a) Energy slack variables. (b) Temperature slack variables.}
    \label{FIG_SIM_4}
\end{figure}

We further assess the roles of the slack variables and balancing bounds. Fig.~\ref{FIG_SIM_4} depicts the evolution of the slack variables over clusters $\sum_{j=1}^k\bar{\xi}_j^{(E)}$ and $\sum_{j=1}^k\bar{\xi}_j^{(T)}$. Note that the cells and the resulting clusters are not fully bounded by the balancing constraints in \eqref{ConvexBalancingConstraints}. Because of this, the power optimization problem in \eqref{InterClusterOptimConvex} would have been infeasible to solve even at the beginning time instant without the introduction of the slack variables. However, the slack variables help prevent the infeasibility issue from occurring. They are initially nonzero to slightly relax the balancing constraints and then penalized to decrease toward zero as the balancing constraints are increasingly satisfied. 

\begin{figure}[!t]
    \centering
    \subfloat[\centering ]{{\includegraphics[trim={2.4cm 0cm 3.2cm 0cm},clip,width=8.5cm]{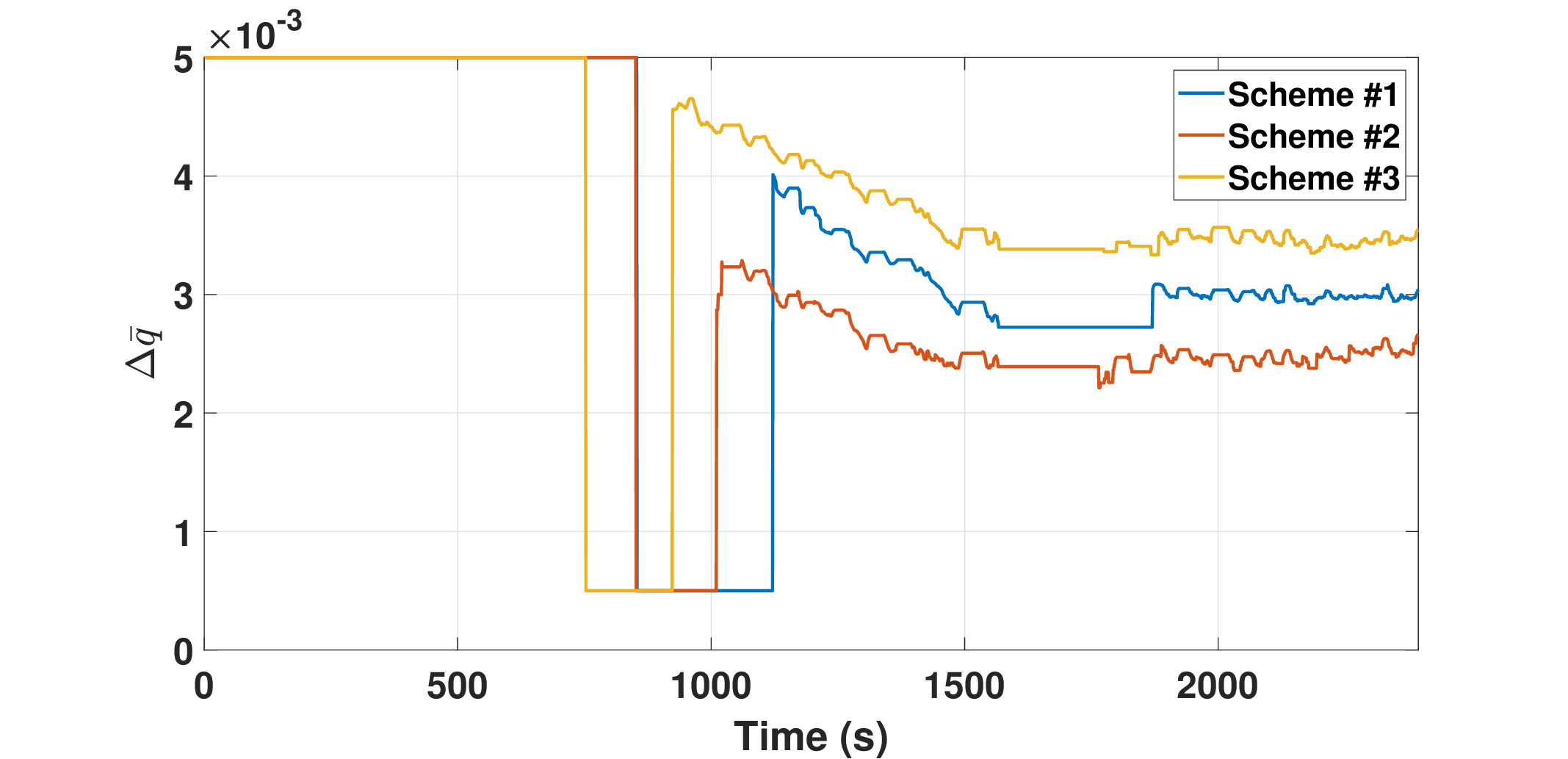} }}
    \;
    \subfloat[\centering ]{{\includegraphics[trim={1.5cm 0cm 3.2cm 1cm},clip,width=8.5cm]{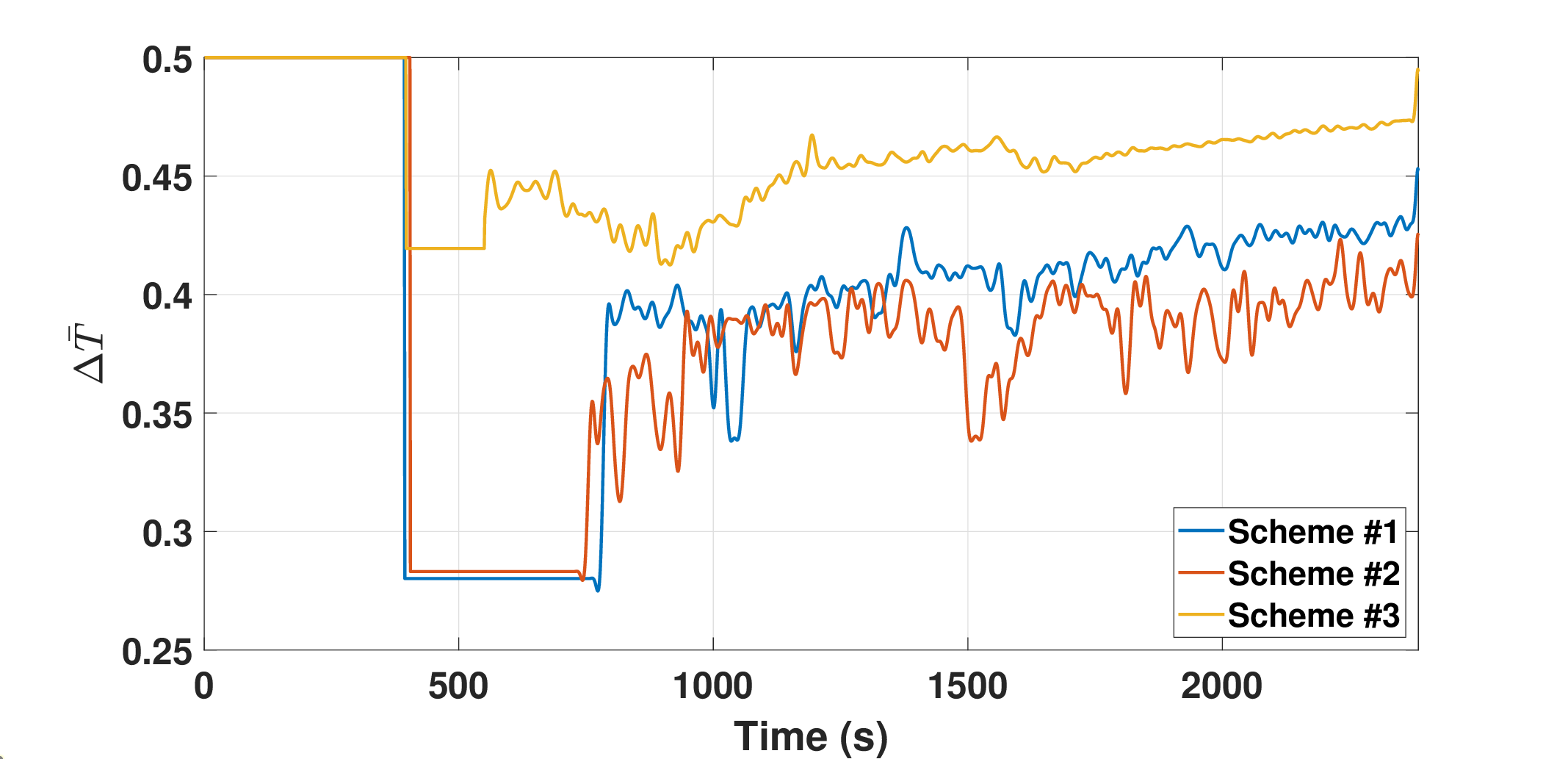} }}
    \;
    \caption{The evolution of the balancing bounds. (a) Energy balancing bounds. (b) Temperature balancing bounds.}
    \label{FIG_SIM_5}
\end{figure}

While zero cluster-level slack variables suggest that the clusters should be balanced, this does not guarantee that all the cells are within the desired bounds, as aforementioned in Section III.C. Fig.~\ref{FIG_SIM_5} shows the utility of the adaptive cell balancing bounds $\Delta \bar{q}$ and $\Delta \bar{T}$. We can see that the original SoC and temperature balancing bounds are 0.5\% and 0.5 K, respectively. According to Fig.~\ref{FIG_SIM_4} (b), the temperature slack variables reach zero at the time instant of 400 seconds. At this moment, the temperature balancing bounds are tightened as shown in Fig.~\ref{FIG_SIM_5} (b), which leads to an increase in the temperature slack variables. Fig.~\ref{FIG_SIM_4} (a) also indicates that the SoC slack variables reach zero between the time steps of 750-900 seconds for the three power split schemes. At the same time instants, we can see an update in the SoC balancing bounds in Fig.~\ref{FIG_SIM_5} (a). We can conclude from the evolution of the balancing bounds the usefulness of tighter bounds for the cluster-level balancing to improve the uniformity at the cell level.  

\begin{figure}[!t]
\centering
\includegraphics[trim={0 0 0 0cm},clip, width=8.5cm]{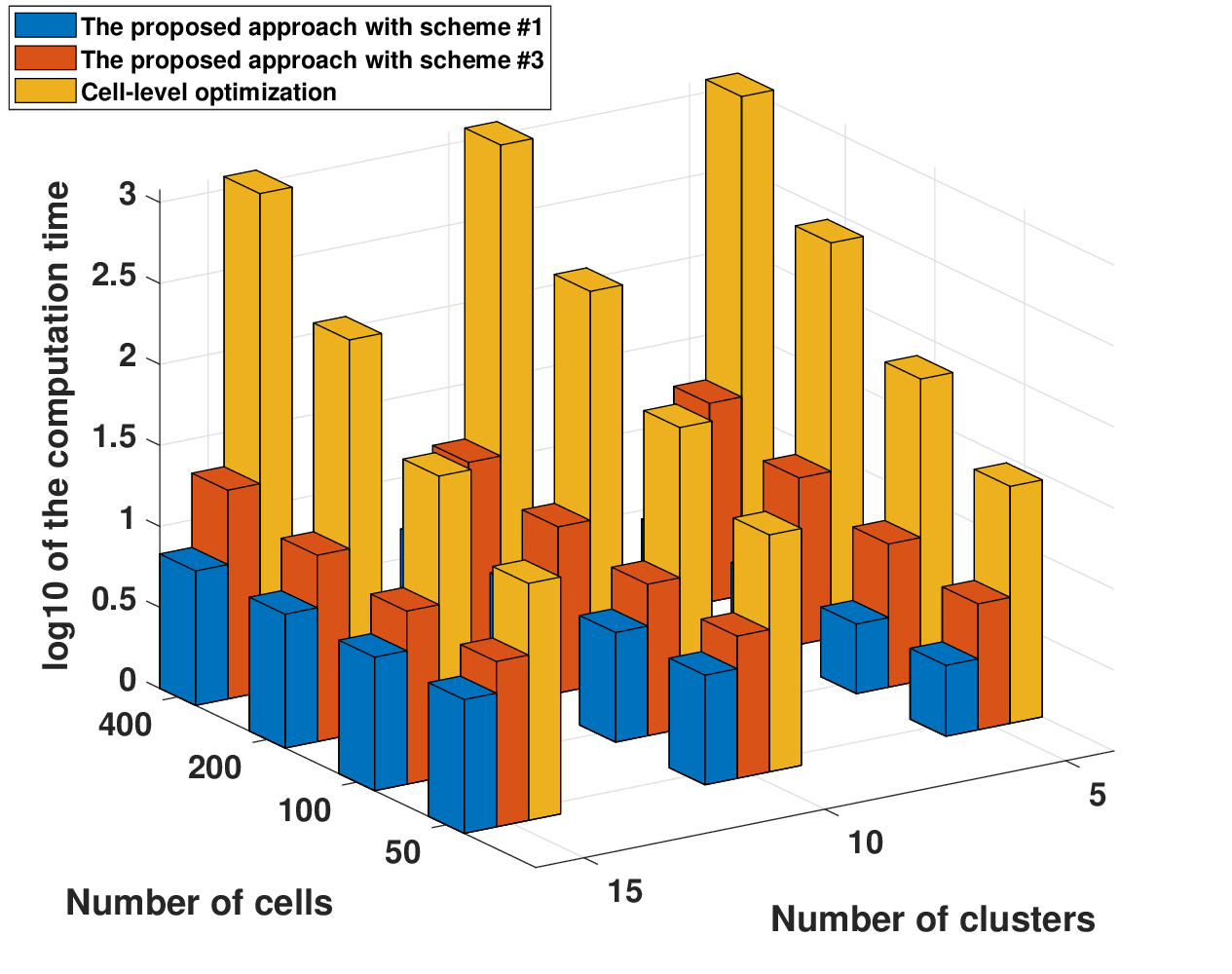}
\caption{Comparative analysis of computation times: the proposed approach with schemes \#1 and \#3 versus the cell-level optimization.}
\label{Computation} 
\end{figure}

As a main objective of the study is to enable computationally efficient optimal power management, we measure the computation time of the approach under schemes \#1 and \#3 for different cell numbers and cluster numbers. Note that scheme \#2 is skipped as its computational performance is similar to that of scheme \#1. We also compare the proposed approach with cell-based power optimization as in the works \cite{TTE-FA-2022, 2019-TVT-CR, 2016-TSTE-PC}. Table \ref{Table: Computation-Comp} summarizes the computation time, with further illustration shown in Fig. \ref{Computation}. We observe several key findings. Firstly, the computation time for scheme \#1 remains independent of the cell number and depends only on the number of clusters. Secondly, the proposed approach under scheme \#3 requires more computation time than scheme \#1 due to the optimal power split among the constituent cells of the clusters. Finally, our proposed approach exhibits substantially less computation time compared to the cell-level optimization. The improvement is more significant when dealing with larger numbers of cells. For instance, our proposed approach takes 6.76 and 19.43 seconds for schemes \#1 and \#3 with 15 clusters, less than 2\% of the about 1200 seconds required by the cell-level optimization. The results validate the computational efficiency and scalability of the proposed approach for large-scale BESS.

\begin{table}[t!]\centering
\caption{Numerical comparison of the proposed approach and cell-level optimization}
\resizebox{0.48\textwidth}{!}{
 \begin{tabular}{>{\color{black}}c | >{\color{black}}l  >{\color{black}}c  >{\color{black}}c}
\toprule
\makecell[c]{Cell Number ($n$)} & \makecell[c]{Method} & \makecell[c]{ Average \\ Computation \\ Time (s)} & \makecell[c]{Relative \\ Computation \\ Time Reduction\\ (\%)}  \\
\midrule
\multirow{7}{0.5cm}{\makecell[c]{50}} & {\bf Cell-level optimization} & 29.02 & --- \\
& {\bf Scheme \#1} (15 clusters) & 6.69 & 76.94  \\
& {\bf Scheme \#1} (10 clusters) & 4.75 & 83.63  \\
& {\bf Scheme \#1} (5 clusters) & 2.74 & 90.55  \\
\cmidrule(l){2-4}
& {\bf Scheme \#3} (15 clusters) & 10.49 & 63.85  \\
& {\bf Scheme \#3} (10 clusters) & 7.55 & 73.98  \\
& {\bf Scheme \#3} (5 clusters) & 5.99 & 79.35  \\
\cmidrule(l){1-4}

\multirow{7}{2cm}{\makecell[c]{100}} & {\bf Cell-level optimization} & 72.77 & --- \\
& {\bf Scheme \#1} (15 clusters) & 6.67 & 90.83  \\
& {\bf Scheme \#1} (10 clusters) & 4.75 & 93.47  \\
& {\bf Scheme \#1} (5 clusters) & 2.69 & 96.30  \\
\cmidrule(l){2-4}
& {\bf Scheme \#3} (15 clusters) & 11.71 & 83.90  \\
& {\bf Scheme \#3} (10 clusters) & 8.61 & 88.16 \\
& {\bf Scheme \#3} (5 clusters) & 7.65 & 89.48 \\
\cmidrule(l){1-4}

\multirow{7}{0.5cm}{\makecell[c]{200}} & {\bf Cell-level optimization} & 274.97 & --- \\
& {\bf Scheme \#1} (15 clusters) & 6.70 & 97.56  \\
& {\bf Scheme \#1} (10 clusters) & 4.75 & 98.27  \\
& {\bf Scheme \#1} (5 clusters) & 2.76 & 98.99  \\
\cmidrule(l){2-4}
& {\bf Scheme \#3} (15 clusters) & 14.13 & 94.86  \\
& {\bf Scheme \#3} (10 clusters) & 10.61 & 96.14  \\
& {\bf Scheme \#3} (5 clusters) & 10.68 & 96.11  \\
\cmidrule(l){1-4}

\multirow{7}{0.5cm}{\makecell[c]{400}} & {\bf Cell-level optimization} & 1200.45 & --- \\
& {\bf Scheme \#1} (15 clusters) & 6.76 & 99.43  \\
& {\bf Scheme \#1} (10 clusters) & 4.83 & 99.59  \\
& {\bf Scheme \#1} (5 clusters) & 2.79 & 99.76  \\
\cmidrule(l){2-4}
& {\bf Scheme \#3} (15 clusters) & 19.43 & 98.38  \\
& {\bf Scheme \#3} (10 clusters) & 14.46 & 98.79  \\
& {\bf Scheme \#3} (5 clusters) & 16.80 & 98.60  \\
\cmidrule(l){1-4}
\end{tabular}
 }
\label{Table: Computation-Comp}
\end{table}

\section{Experimental Results}
\begin{figure}[!t]
    \centering
    \subfloat[\centering ]{{\includegraphics[trim={0cm 1.5cm 0cm 0cm},clip,width=\linewidth]{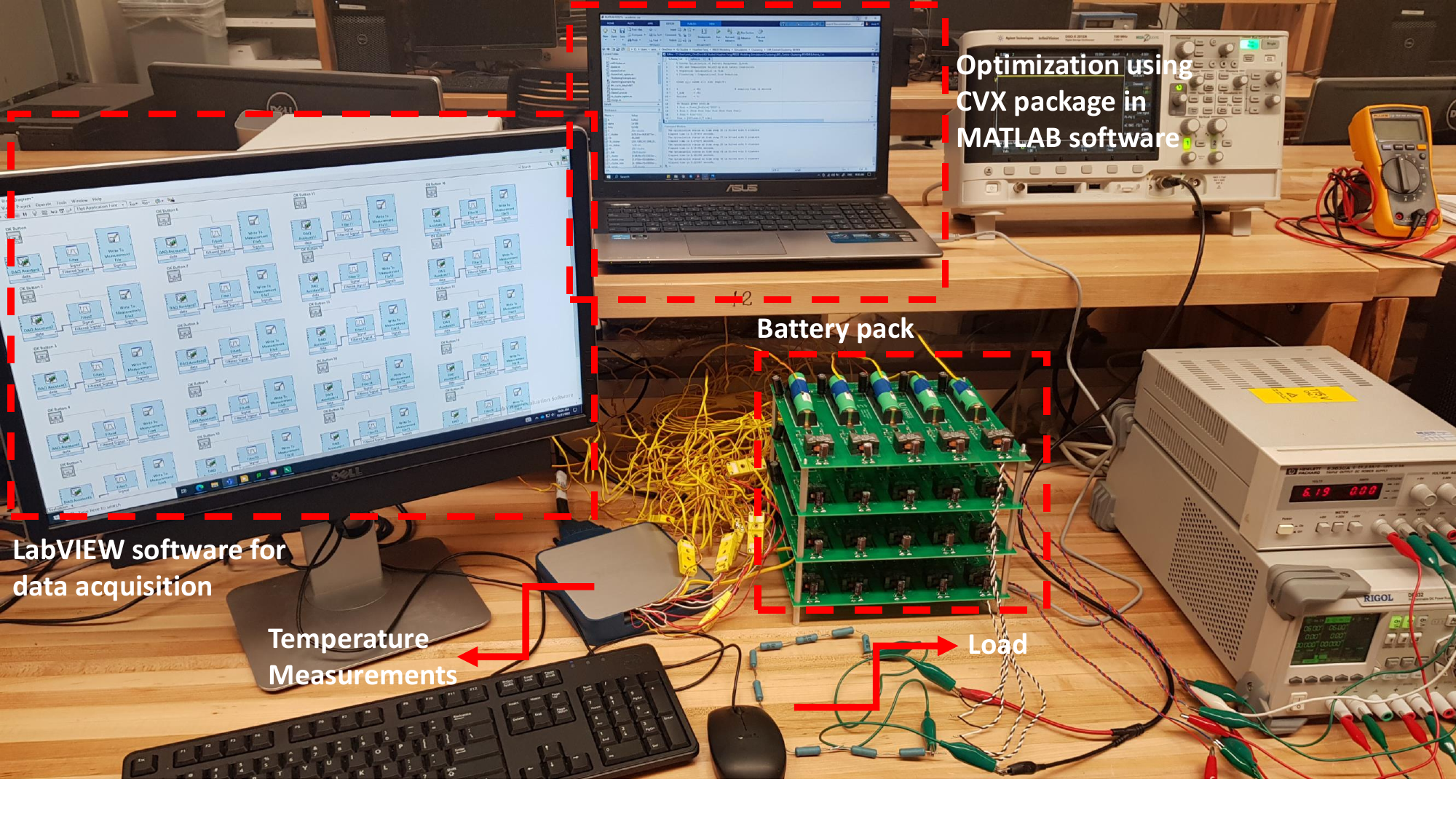} }}
    \;
   \vspace{0.5em} \subfloat[\centering ]{{\includegraphics[trim={0cm 0cm 0cm 0cm},clip,width=5cm]{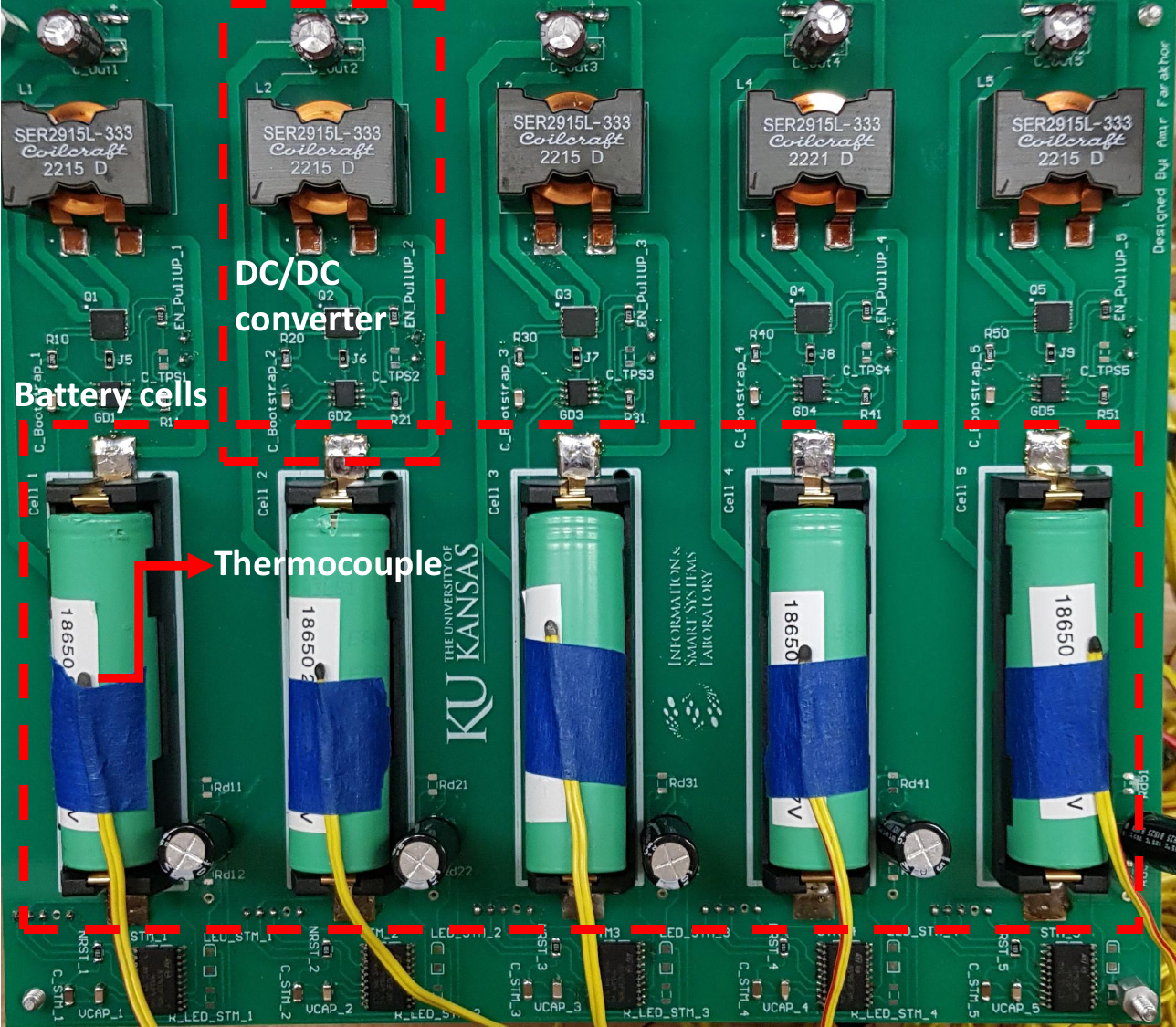} }}
    \;
    \caption{The lab-scale prototype. (a) The experimental setup. (b) The circuit.}
    \label{FIG_EXP_1}
\end{figure}

We develop a lab-scale prototype to validate the proposed optimal power management approach. The experimental validation encompasses various aspects, including the evaluation of cell clustering, model aggregation, optimization, and cluster-to-cell power split. Fig.~\ref{FIG_EXP_1} (a) depicts the experimental setup, and Fig.~\ref{FIG_EXP_1} (b) shows a 20-cell 4s5p battery pack. Table \ref{TABLE_2} lists the specifications of the key components of the battery pack. We use K-type thermocouples to measure the surface temperatures of the cells. The thermocouples are connected to a National Instruments PCIe-6321 DAQ board, and the measurements are collected via LabVIEW. We solve the optimization problem using the CVX package \cite{CVX-1,CVX-2}, and the optimal control decisions are transferred to local controllers through DSP TMS320F28335. The local controllers generate 200-kHz switching signals for the DC/DC converters. We use a 150-W power load to discharge the battery pack. The experiment lasts thirty minutes with a step size of thirty seconds for the optimization, i.e., $\Delta t = 30$ s.

\begin{figure*}[t]
    \centering
    \subfloat[\centering ]{{\includegraphics[trim={1.8cm 0 2.5cm 1cm},clip,width=4.3cm]{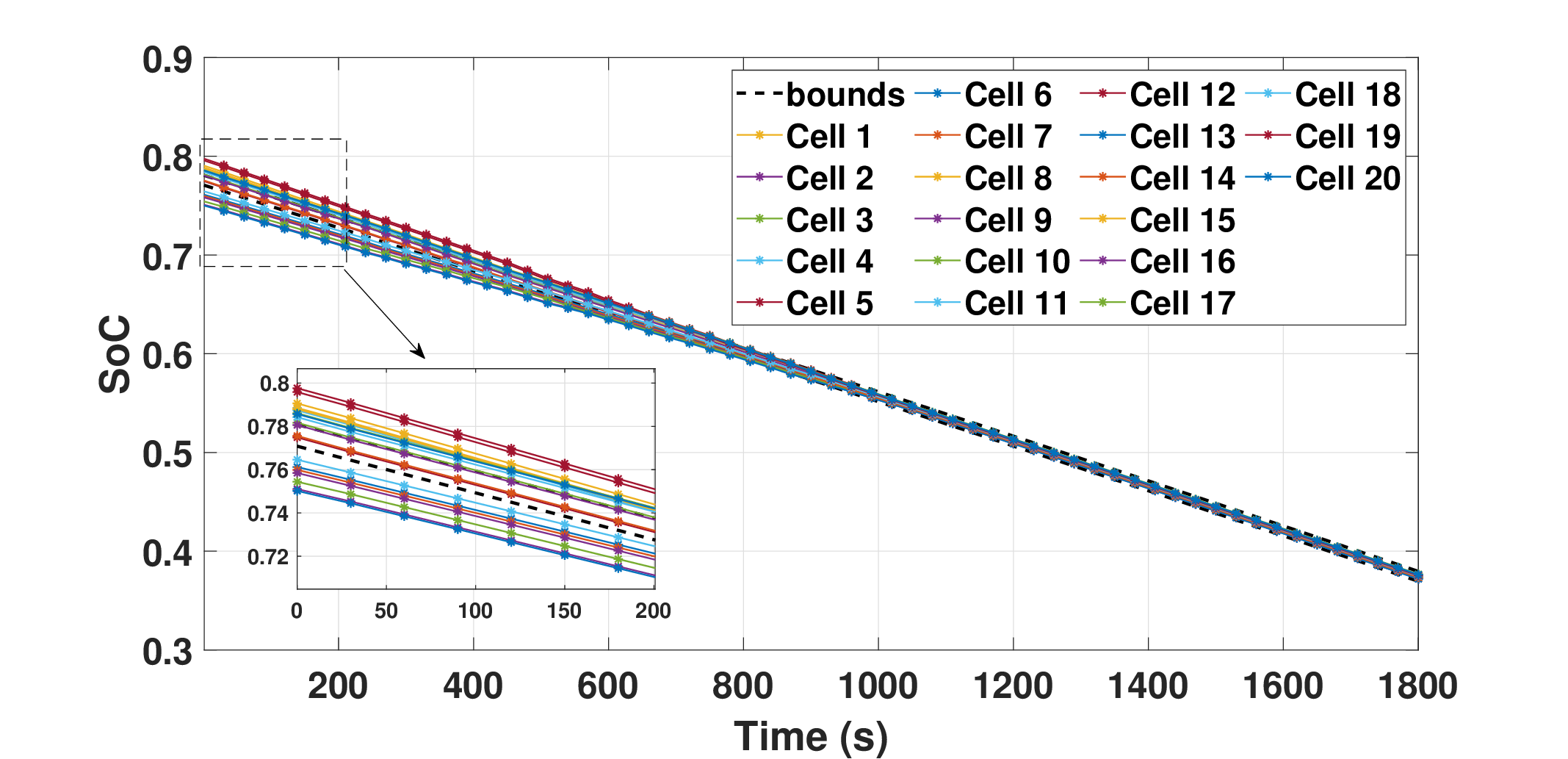} }}
    \subfloat[\centering ]{{\includegraphics[trim={1.8cm 0 2.5cm 1cm},clip,width=4.3cm]{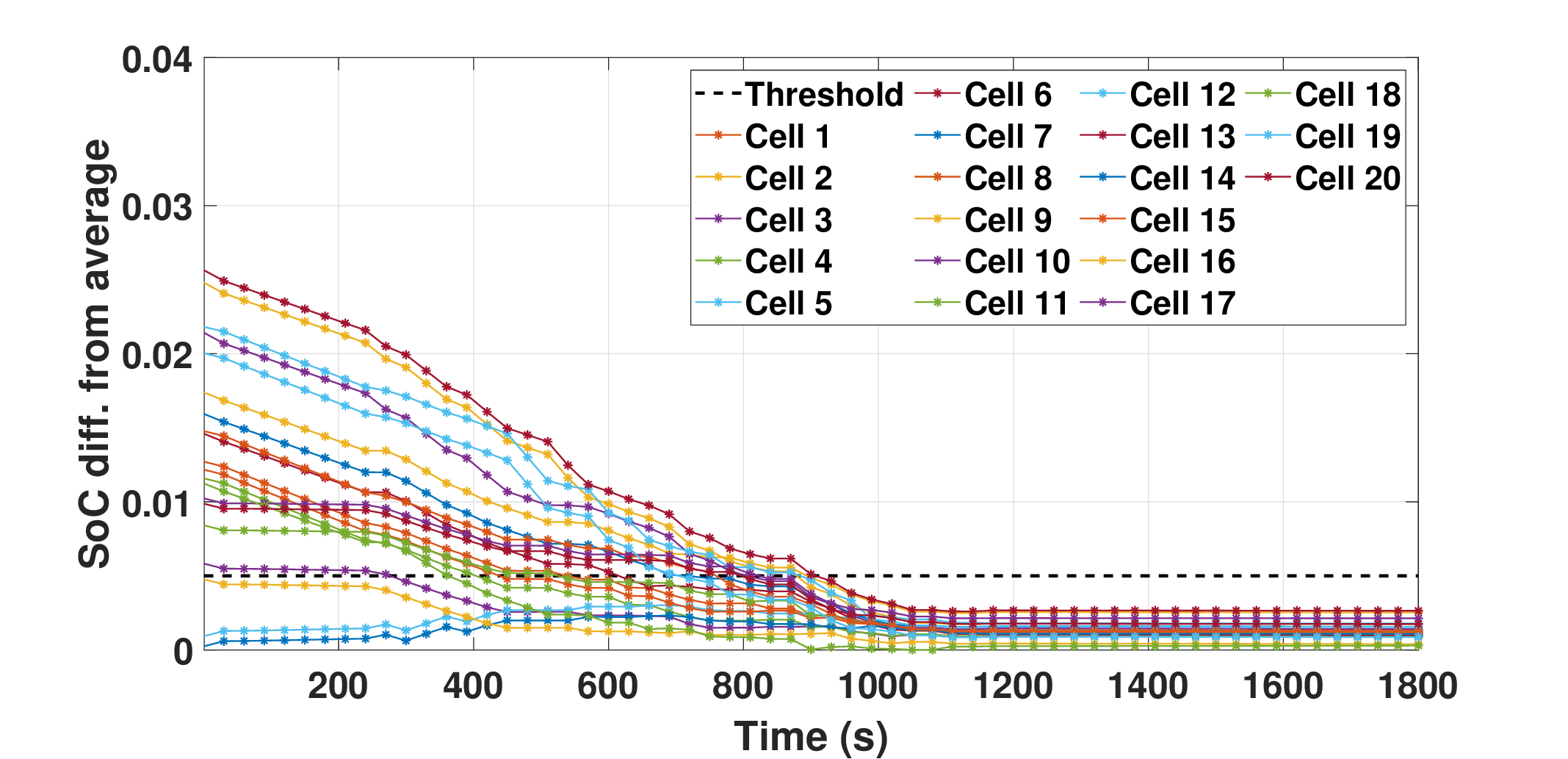} }}
    \subfloat[\centering ]{{\includegraphics[trim={1.8cm 0 2.5cm 1cm},clip,width=4.3cm]{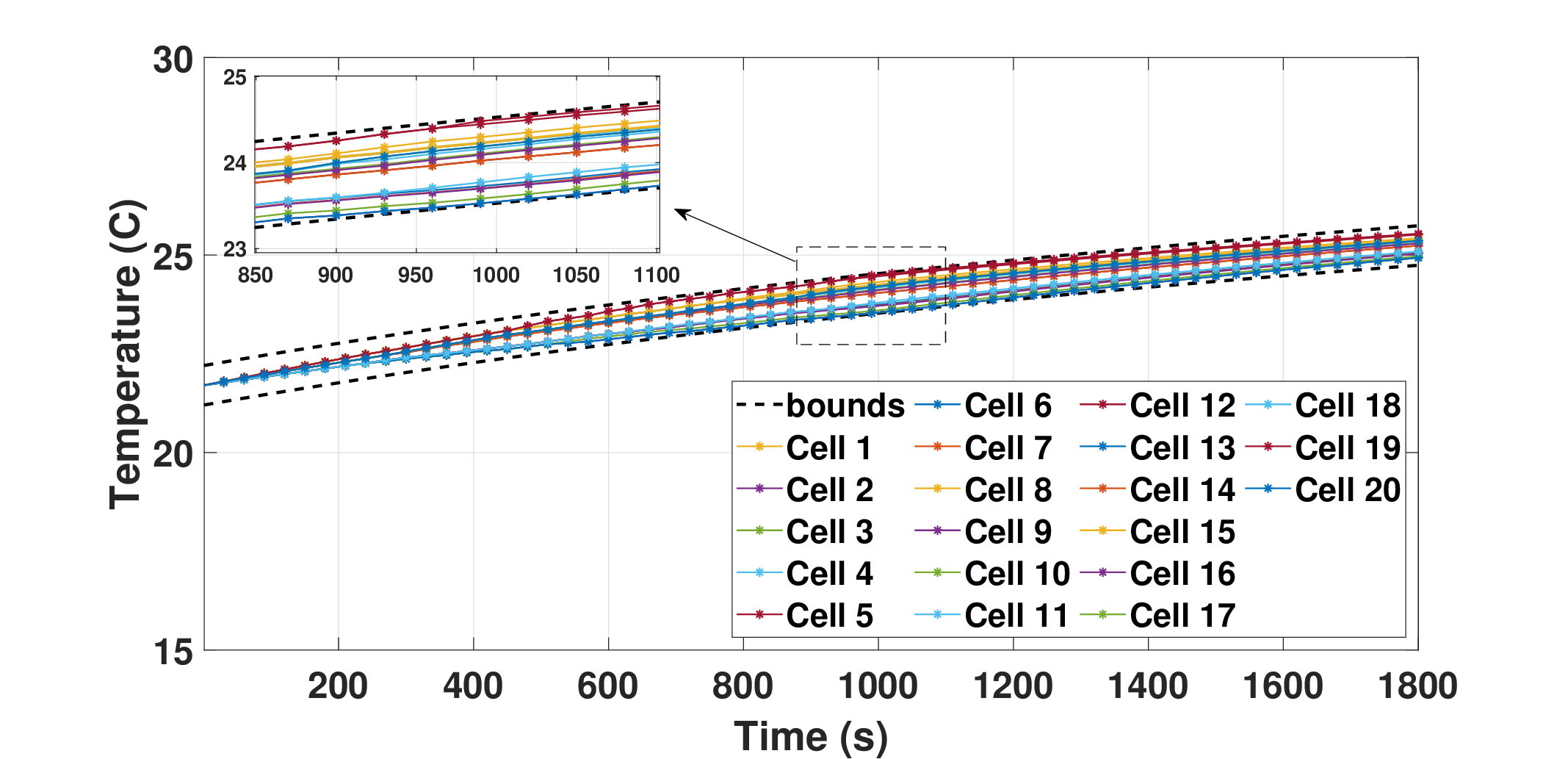} }}
    \subfloat[\centering ]{{\includegraphics[trim={1.5cm 0 2.5cm 1cm},clip,width=4.3cm]{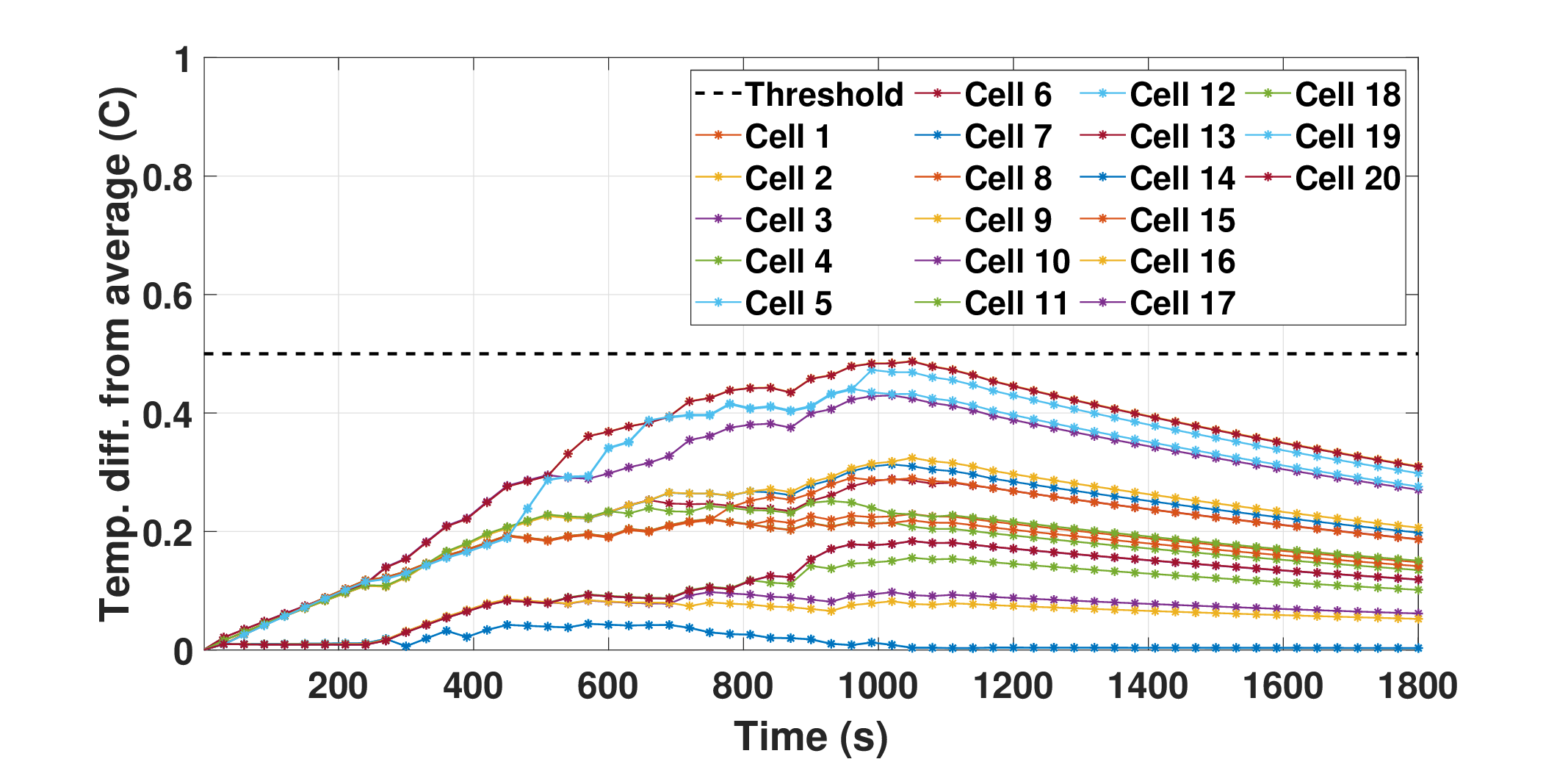} }}

    \subfloat[\centering ]{{\includegraphics[trim={1.8cm 0 2.5cm 1cm},clip,width=4.3cm]{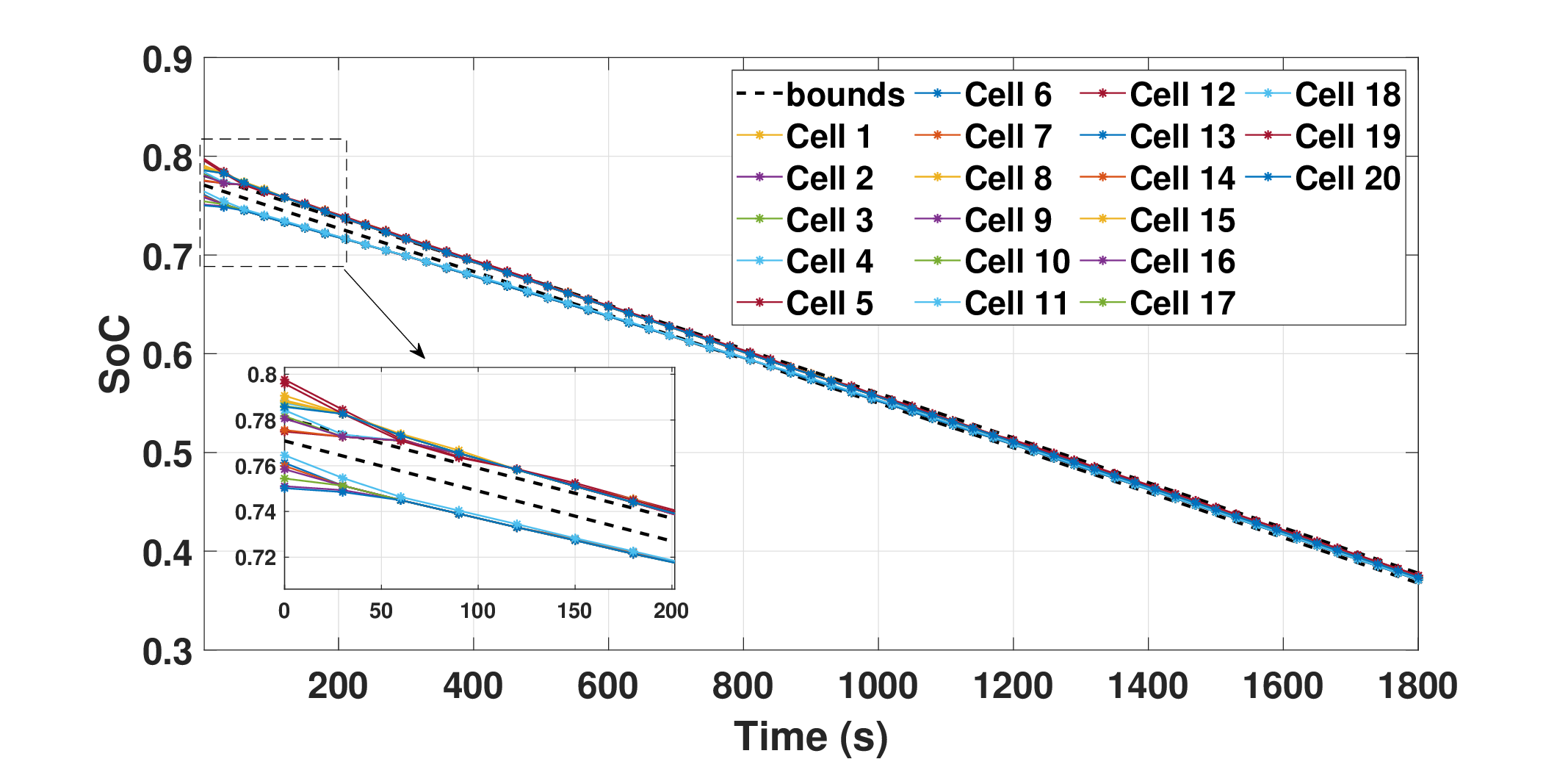} }}
    \subfloat[\centering ]{{\includegraphics[trim={1.8cm 0 2.5cm 1cm},clip,width=4.3cm]{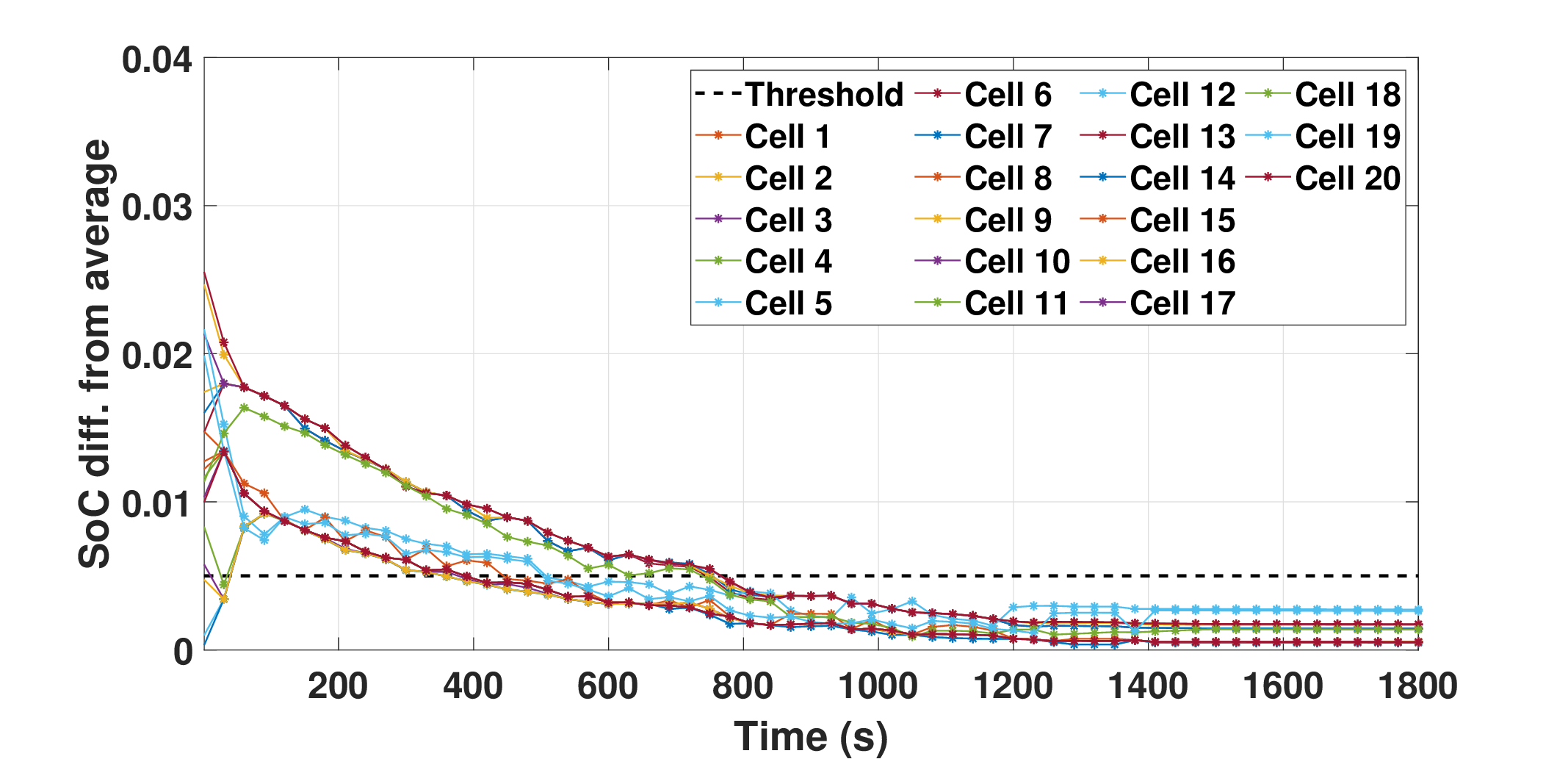} }}
    \subfloat[\centering ]{{\includegraphics[trim={1.8cm 0 2.5cm 1cm},clip,width=4.3cm]{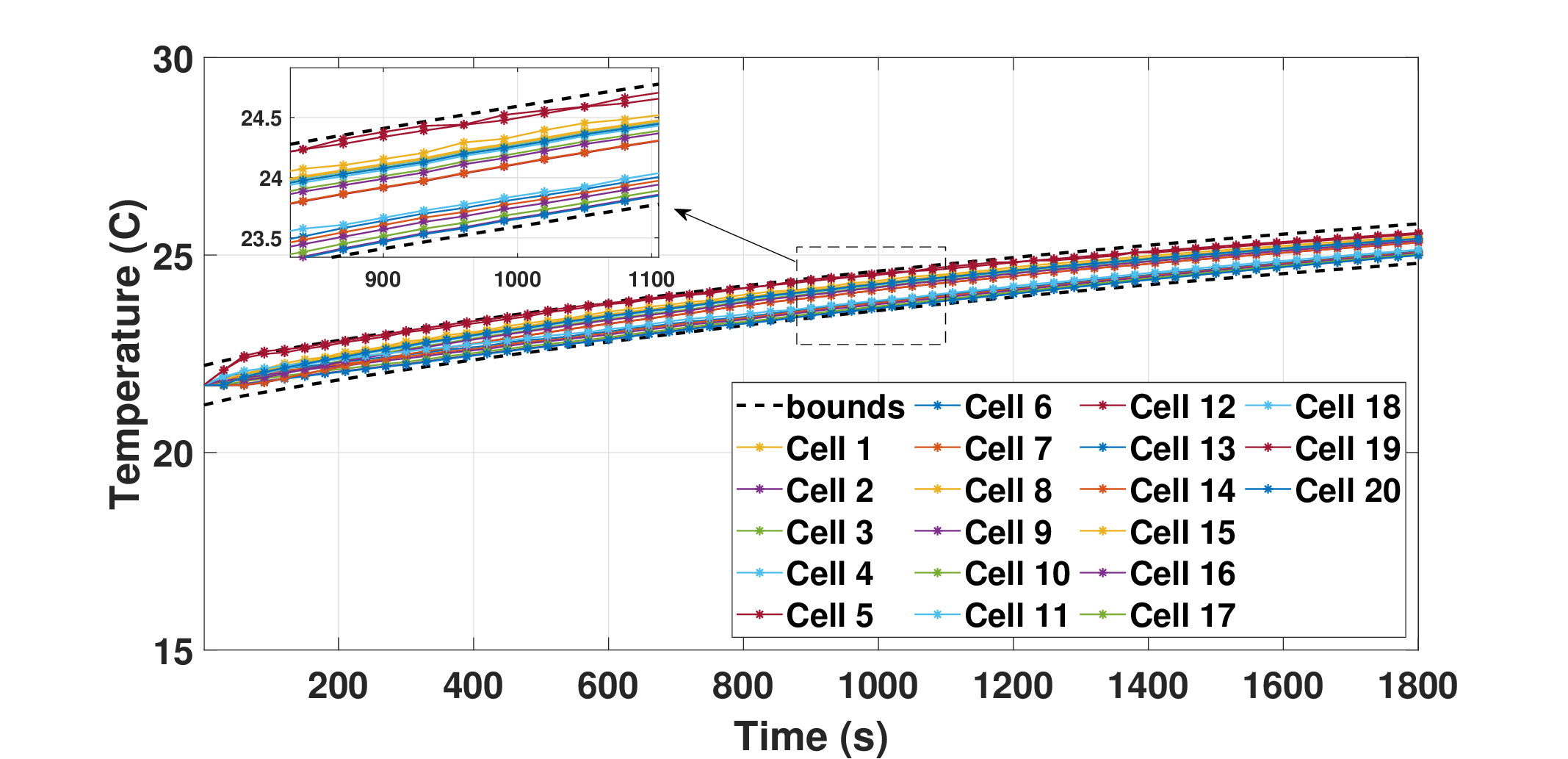} }}
    \subfloat[\centering ]{{\includegraphics[trim={1.5cm 0 2.5cm 1cm},clip,width=4.3cm]{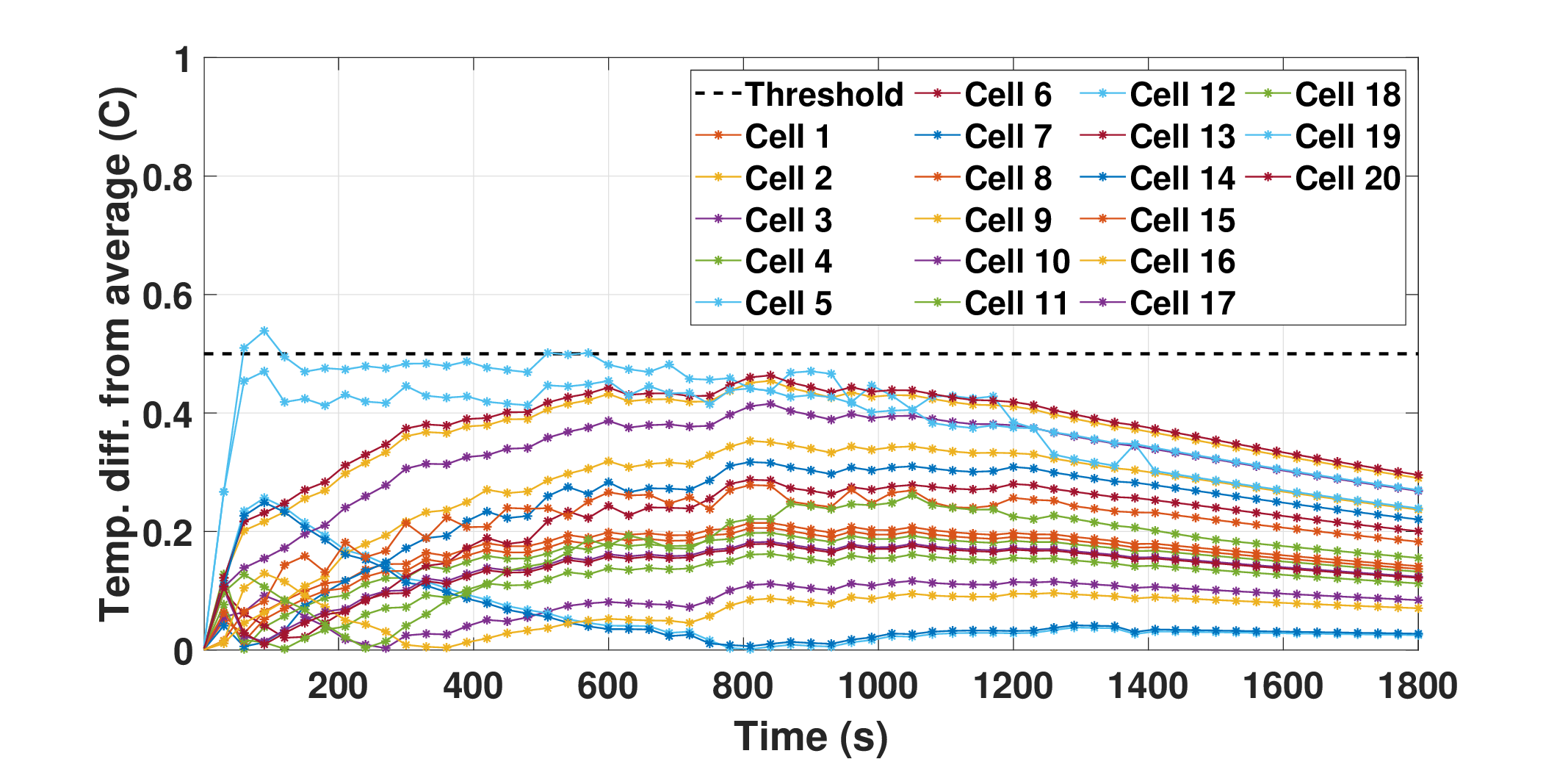} }}

    \caption{Experimental results of the SoC and temperature balancing. (a) The SoC of the cells in scheme \#1. (b) The SoC difference of the cells from the average in scheme \#1. (c) The temperature of the cells in scheme \#1. (d) The temperature difference of the cells from the average in scheme \#1. (e) The SoC of the cells in scheme \#3. (f) The SoC difference of the cells from the average in scheme \#3. (g) The temperature of the cells in scheme \#3. (h) The temperature difference of the cells from the average in scheme \#3.}
    \label{FIG_EXP_2}
\end{figure*}

\begin{figure}[!t]
    \centering
    \subfloat[\centering ]{{\includegraphics[trim={1.8cm 0 2cm 1cm},clip,width=8.5cm]{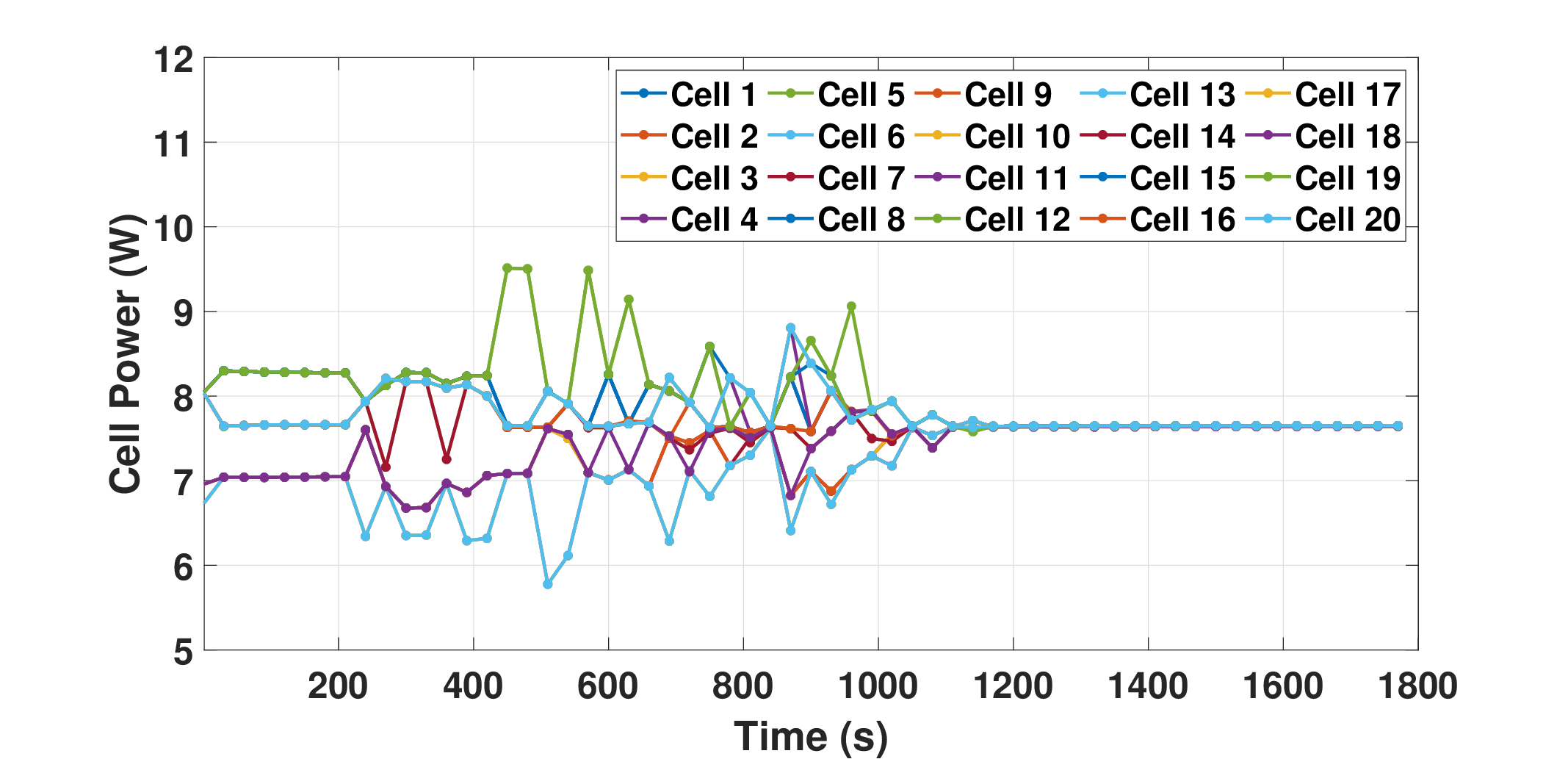} }}
    \;
    \subfloat[\centering ]{{\includegraphics[trim={1.8cm 0 2cm 1cm},clip,width=8.5cm]{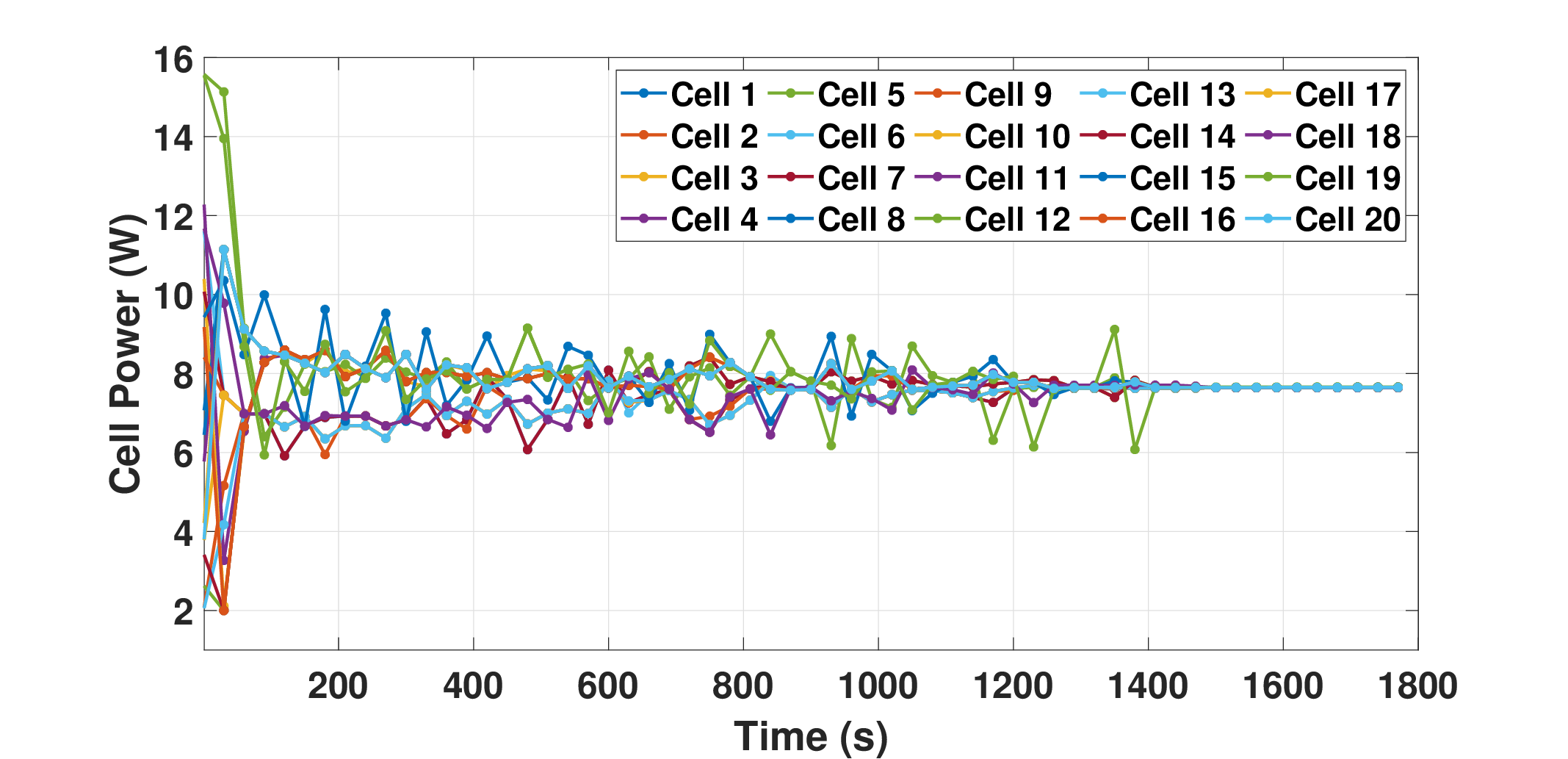} }}
    \;
    \caption{The output power profiles of the cells. (a) Scheme \#1. (b) Scheme \#3.}
    \label{FIG_EXP_3}
\end{figure}

The cells' initial SoC ranges from 75\% to 80\%, whereas their initial temperature is 21.7\degree C. The cells' output discharging current is limited to be no more than 5 A. In the experiment, we also examine the performance of the three cluster-to-cell power split schemes. Because the constituent cells are identical, scheme \#2 will reduce to scheme \#1, and both will generate the same results. Therefore, we will apply only schemes \#1 and \#3 to do the power split in the experiment. The experimental results are given in Figs.~\ref{FIG_EXP_2}-\ref{FIG_EXP_3}.

\begin{table}[!b]
	\renewcommand{\arraystretch}{1.3}
	\caption{List of Key Components in the Experimental Setup}
	\centering
	\label{TABLE_2}
	\centering
	{
		\begin{tabular}{l l}
			\hline\hline \\[-3mm]
			\multicolumn{1}{c}{Device} & \multicolumn{1}{c}{Model (Value)}\\[1.6ex] \hline
			MOSFET & CSD86356Q5D\\
			Gate driver & TPS28225\\
			Inductor & SER2915H-333KL (33 $\mu$H) \qquad \qquad \\
			Capacitor & (10 $\mu$F)\\
			Local controller \qquad \qquad \qquad & STM8S003F3P6\\
			Main controller & TMS320f28335\\
			Battery cell & Samsung INR18650-25R\\
			\hline\hline
		\end{tabular}
	}
\end{table}

Fig.~\ref{FIG_EXP_2} (a) shows the cells' SoC profiles under scheme \#1. We can observe that the initial SoC values of the cells are not balanced and do not lie within the desired bound. However, the proposed approach successfully distributes the output power among different cells to drive the cells' SoC to enter the tolerable bounds. Fig.~\ref{FIG_EXP_2} (b) illustrates the SoC difference of the cells from the average for a better visualization of the SoC balancing trend. It can be seen that scheme \#1 manages to balance cells' SoC after about 900 seconds. Fig.~\ref{FIG_EXP_2} (c) and Fig.~\ref{FIG_EXP_2} (d) depict the cells' temperature profiles and the differences from the average, respectively. The cells see their temperatures drift away from each other due to the uneven power allocation among the cells, but the deviation is bounded within the tolerance range. Note that, under scheme \#1, the cells are categorized to three to six clusters during the experiment. 

Fig.~\ref{FIG_EXP_2} (e) and Fig.~\ref{FIG_EXP_2} (f) show the cells' SoC profiles and the deviation from the average under scheme \#3. In this case, the SoC is balanced among the cells after about 750 seconds, faster than when scheme \#1 is applied, because of the optimal intra-cluster power split. Fig.~\ref{FIG_EXP_2} (g) and Fig.~\ref{FIG_EXP_2} (h) also depict the temperature of the cells and the temperature difference from the average, respectively. Similar to Figs.~\ref{FIG_EXP_2} (c)-(d), the cells' temperatures diverge mildly from the same initial point, but the divergence is almost constantly bounded throughout the experiment. There is a minor violation in the temperature balancing constraint around 70-80 seconds which is allowed by the slack variables for the sake of the feasibility of the optimization problem. This experiment finds a maximum of four clusters, fewer than under scheme \#1.

Figs.~\ref{FIG_EXP_3} (a)-(b) illustrate the output profiles of the cells under schemes \#1 and \#3, respectively. Both show the cells are assigned different power levels based on their conditions, and after the balancing is achieved eventually, they are assigned the same power. Yet, there is an interesting difference in Figs.~\ref{FIG_EXP_3} (a)-(b). Scheme \#1 distributes the total power load equally among the cells within a cluster; by contrast, scheme \#3 performs intra-cluster power split in an optimal way, leading to more variability in power allocation, but faster adjustment, among the cells. 

\section{Conclusion and Future Work}
The rapidly  growing use of large-scale BESS is transforming grid,  transportation, and more sectors towards decarbonization and a  sustainable future. However, an open challenge for this technology is optimal power management, which struggles with high computational complexity in optimization. Thus, this paper proposes a computationally efficient and scalable approach to deal with the challenge. The approach distinguishes itself from the literature by enabling clustering-based optimization. It partitions a large number of cells into just a much smaller number of clusters based on their characteristics and develops aggregated electro-thermal models for the clusters. Inter-cluster power management then naturally arises to compute the power quota among the clusters at considerably high computational efficiency. Intra-cluster power split then divides the power quota among the cells within each cluster under three different schemes. We further propose adaptive cell balancing bounds to improve the cell-level control performance. The proposed approach is designed to perform cell balancing in terms of SoC and temperature while minimizing the total power losses. We conduct simulations for a 400-cell BESS to validate the computational efficiency and scalability of the proposed approach. The results show a substantial reduction in computational overhead, by margins as large as 98\% for a 400-cell BESS when compared to conventional cell-level optimization. Further, we develop a lab-scale 20-cell prototype for practical validation. The results show that the approach can effectively reduce power losses, enhance cell balancing, and require much less computational costs. The study can be expanded in several directions. Our future work will include extending the proposed approach to second-life battery systems, incorporating more sophisticated battery models, and studying multi-level clustering for stronger scalability.

\bibliographystyle{IEEEtran}
\scriptsize\bibliography{IEEEabrv,Bibliography/BIB}

\begin{IEEEbiography}[{\includegraphics[width=1in,height=1.25in,clip,keepaspectratio]{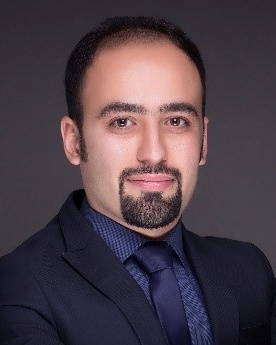}}]
{Amir Farakhor} (Graduate Student Member, IEEE) received the B.Sc. and M.Sc. degrees in electrical engineering from the Azarbaijan Shahid Madani University, Tabriz, Iran in 2012 and 2014, respectively, and Ph.D. degree in Power Electronics from the University of Tabriz in Feb 2019. He is currently a Ph.D. candidate in mechanical engineering at the University of Kansas, Lawrence, KS, USA. His research interests include power electronics, battery management systems, renewable energies, and distributed generation.
\end{IEEEbiography}

\begin{IEEEbiography}[{\includegraphics[width=1in,height=1.25in,clip,keepaspectratio]{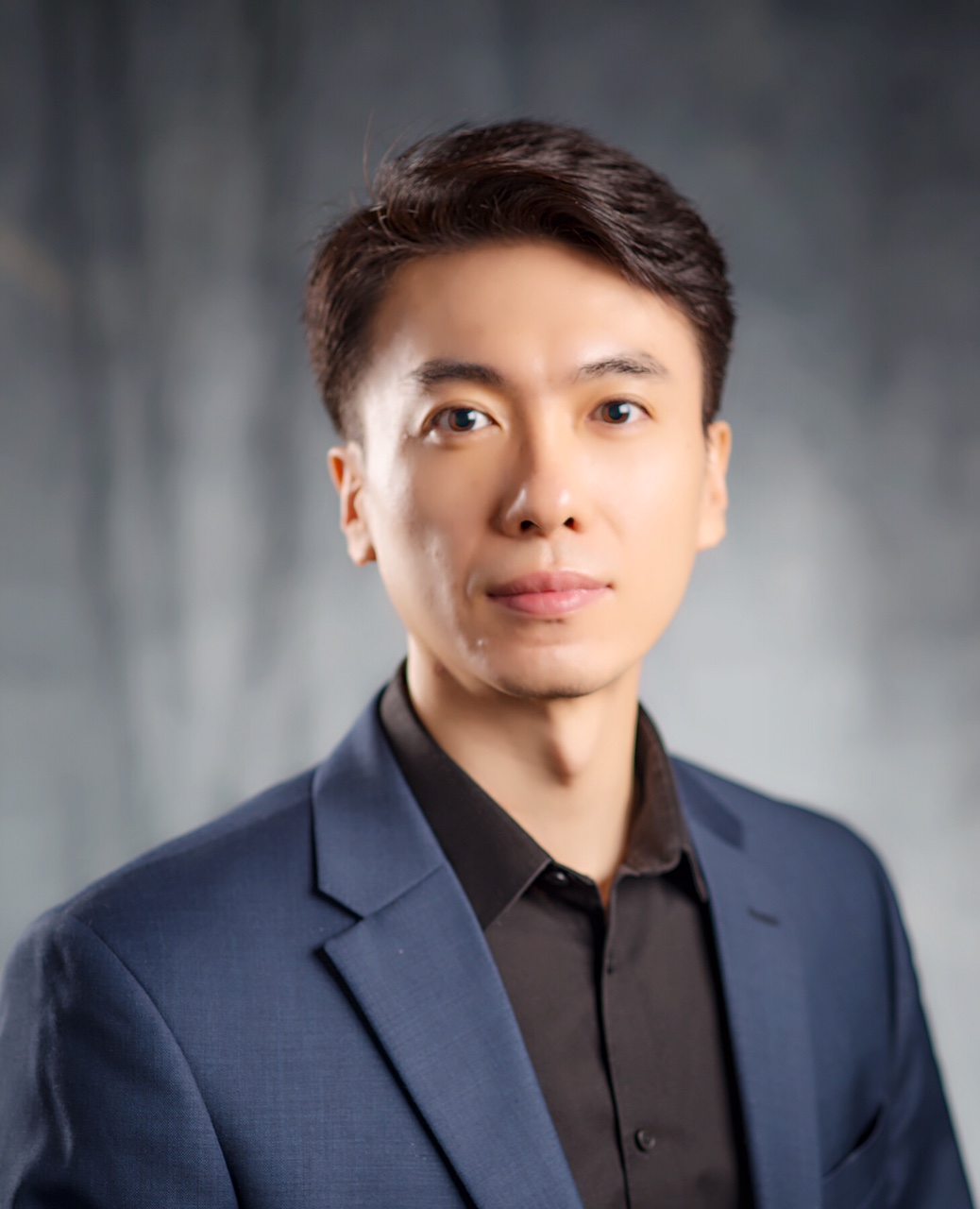}}]
{Di Wu} (Senior Member, IEEE) is a Chief Research Engineer and a Team Leader within the Optimization and Control Group at the Pacific Northwest National Laboratory (PNNL). He received the B.S. and M.S. degrees in electrical engineering from Shanghai Jiao Tong University, China, in 2003 and 2006, respectively, and the Ph.D. in electrical and computer engineering from Iowa State University, Ames, in 2012. At PNNL, Dr. Wu leads research work in the areas of energy storage analytics, building-to-grid integration, microgrid design, and hybrid energy systems. Dr. Wu is a Senior Member of IEEE and a member of the IEEE Power and Energy Society and the Control System Society. He serves as an Editor for the IEEE Open Access Journal of Power and Energy and IEEE Transactions on Energy Markets, Policy and Regulation.
\end{IEEEbiography}

\begin{IEEEbiography}[{\includegraphics[width=1in,height=1.25in,clip,keepaspectratio]{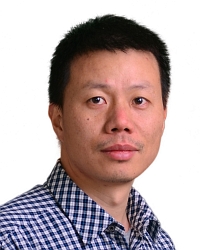}}]
{Yebin Wang} (Senior Member, IEEE) received the B.Eng. degree in mechatronics engineering
from Zhejiang University, Hangzhou, China, in 1997, the M.Eng. degree in control theory and control engineering from Tsinghua University, Beijing, China, in 2001, and the Ph.D. degree in electrical engineering from the University of Alberta, Edmonton, AB, Canada, in 2008.
He has been with Mitsubishi Electric Research Laboratories, Cambridge, MA, USA, since 2009, where he is currently a Senior Principal Research Scientist. From 2001 to 2003, he was a Software Engineer, the Project Manager, and the Manager of the Research and Development Department in Industries, Beijing, China. His current research interests include nonlinear control and estimation, optimal control, adaptive systems, and their applications, including mechatronic systems.
\end{IEEEbiography}

\begin{IEEEbiography}[{\includegraphics[width=1in,height=1.25in,clip,keepaspectratio]{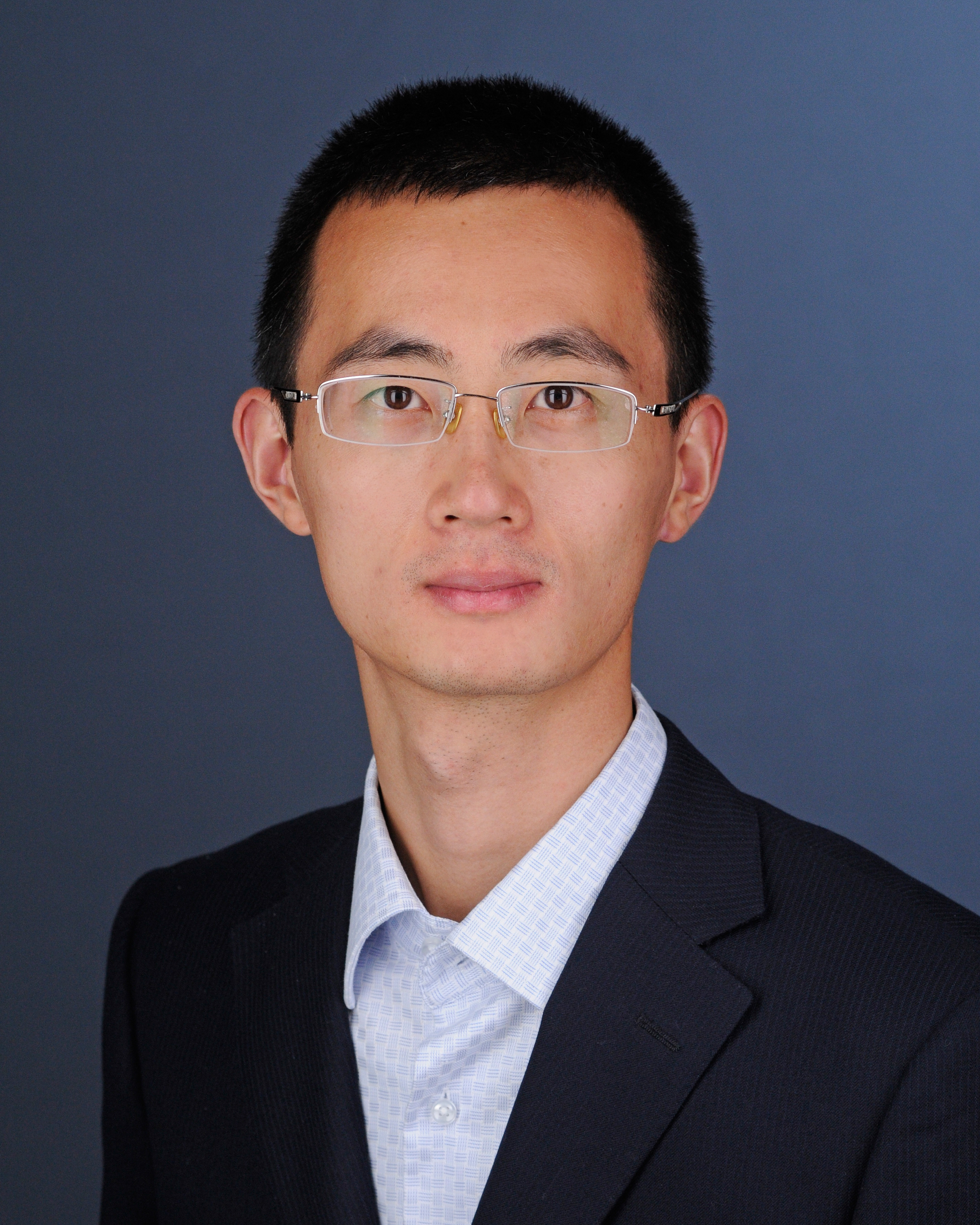}}]
{Huazhen Fang} (Member, IEEE) received the B.Eng. degree in computer science and technology from Northwestern Polytechnic University, Xi'an, China, in 2006, the M.Sc. degree in mechanical engineering from the University of Saskatchewan, Saskatoon, Canada, in 2009, and the Ph.D. degree in mechanical engineering from the Department of Mechanical and Aerospace Engineering, University of California, San Diego, La Jolla, CA, USA, in 2014. 
He is an Associate Professor of mechanical engineering with the University of Kansas, Lawrence, KS, USA. His research interests include control and estimation theory with application to energy management and robotics. Dr. Fang received the National Science Foundation CAREER Award in 2019. He currently serves  as an Associate Editor for IEEE Transactions on Industrial Electronics, IEEE Open Journal of the Industrial Electronics Society, IEEE Control Systems Letters, IEEE Open Journal of Control Systems, and Information Sciences.
\end{IEEEbiography}

\end{document}